\documentclass[lettersize,journal]{IEEEtran}
\usepackage{amsmath,amsfonts}
\usepackage{algorithmic}
\usepackage{algorithm}
\usepackage{array}
\usepackage[caption=false,font=normalsize,labelfont=sf,textfont=sf]{subfig}
\usepackage{textcomp}
\usepackage{stfloats}
\usepackage{url}
\usepackage{verbatim}
\usepackage{graphicx}
\usepackage{cite}

\usepackage{xcolor}

\DeclareMathOperator\erf{erf}

\begin{document}
\title{Analytical characterization of RIS-aided terahertz links in the presence of beam misalignment}

\author{Evangelos N. Papasotiriou, Sotiris Droulias, and Angeliki Alexiou,~\IEEEmembership{Member,~IEEE}

\begin{normalsize} 
Department of Digital Systems, University of Piraeus, Piraeus 18534, Greece.
\end{normalsize} \\
}

\maketitle

\begin{abstract}
The terahertz (THz) frequency band has recently attracted considerable attention in wireless communications as potential candidate for providing the necessary high bandwidth for demanding applications. With increasing frequency, however, the communication link becomes more vulnerable to blockage and pathloss increases. While both effects can be mitigated with the judicious utilization of directional beams and Reconfigurable Intelligent Surfaces (RISs), high directivity could potentially increase the probability of undesired misalignment between the beam that is steered by the RIS, and the user. It is therefore crucial to characterize and understand the stochastic behavior of misalignment in RIS-aided THz links. In this work, beam misalignment in RIS-aided links is studied theoretically and analytical models are derived, the validity of which is verified through numerical calculations. It is demonstrated that there is a distinction in the stochastic behavior of misalignment between pointing errors that occur on the steering plane or normally to the steering plane, with direct consequences on the link robustness on misalignment. The analytical models capture the impact of misalignment under these qualitatively different conditions and provide the necessary tools for assessing the stochastic RIS performance with respect to crucial link parameters, such as the transmitter's beam width, the transmitter-RIS distance, the RIS-receiver distance, and the steering angle of the RIS.
\end{abstract}

\begin{IEEEkeywords}
THz, Reconfigurable-Intelligent-Surface, Beam Misalignment
\end{IEEEkeywords}

\section{Introduction}
\label{sec:introduction}
\IEEEPARstart{T}{oday} the fifth-generation $\left(5\text{ G}\right)$ of wireless communications gradually becomes a reality by the worldwide deployment of networks. Without a doubt, the significance of $5\text{ G}$ to meet the ever increasing demands of higher data rates and bandwidth hungry applications is paramount. However, the expected growth of new wireless technologies and the proliferation of the wireless devices in the coming years will lead to shortage of the available bandwidth~\cite{A:Roadmap_to_6G,B:Next_Gen_Wir_THz_Com_Netw}. To overcome this limitation the terahertz (THz) band, which spans the range of ${0.1\text{--}10\text{ THz}}$ is envisioned as a promising enabler for the sixth-generation $\left(6\text{ G}\right)$ of wireless communications~\cite{A:amakawa2021white,Me_Sc_Reps_21}.

\subsection{Related Work}\label{Sec:Related_Work}
The THz wireless transmissions are significantly degraded by the propagation losses, namely the free space attenuation and the molecular absorption loss, due to the atmospheric water vapor~\cite{Me_Sc_Reps_21,jornet2011,J:Kokkoniemi_100_450GHz,Mag:RIS_pathl_alex}. As a consequence to reduce the effect of the propagation attenuation the THz transceivers must be equipped with highly directive high-gain antennas~\cite{B:Next_Gen_Wir_THz_Com_Netw,J:Kokkoniemi_100_450GHz,Mag:Analytical_Performance_Assessment_of_THz_Wireless_Systems,A:On_millimeter_wave_and_THz_mobile_radio_channel_for_smart_rail_mobility}. However, high antenna directivity makes the THz wireless systems prone to the effects of beam misalignment, thus leading to significant performance degradation~\cite{Mag:Analytical_Performance_Assessment_of_THz_Wireless_Systems,C:EGW_Imp_Beam_Misal_RIS_assist_THz_Sys,B:Next_Gen_Wir_THz_Com_Netw,A:Dir_THz_Commun_Sys_for_6G_Fact_Check}. The phenomenon of misalignment can be attributed to a multitude of reasons such as, transceiver hardware impairments~\cite{Mag:Analytical_Performance_Assessment_of_THz_Wireless_Systems,C:PIMRC_2019_turkey,B:Next_Gen_Wir_THz_Com_Netw,Mag:Perf_Anal_of_THz_Wir_Sys_in_the_Pres_of_Ant_Misal_and_PHN,C:Relay-Based_Blockage_and_Antenna_Misalignment_Mitigation_in_THz_Wireless_Communications,A:Joint_Imp_Ph_err_HD_imp_mob_Int_RIS_over_kappa_mu}, phase estimation errors at the various components of the wireless system~\cite{C:EGW_Imp_Beam_Misal_RIS_assist_THz_Sys,C:Eff_Rate_of_RIS_Netw_Loc_and_Ph_Est_Unc,A:Joint_Imp_Ph_err_HD_imp_mob_Int_RIS_over_kappa_mu,C:OP_Anal_of_RIS_Wir_Netw_W_Von_Mises_Ph_Er}, dynamic wind loads and thermal expansions that effect antennas placed on high rise building~\cite{Mag:Analytical_Performance_Assessment_of_THz_Wireless_Systems,C:PIMRC_2019_turkey,B:Next_Gen_Wir_THz_Com_Netw,Mag:Perf_Anal_of_THz_Wir_Sys_in_the_Pres_of_Ant_Misal_and_PHN,C:Relay-Based_Blockage_and_Antenna_Misalignment_Mitigation_in_THz_Wireless_Communications}, etc. In addition, with increasing operation frequency within the THz range, the available wavelength becomes comparable to the size of the objects laid in the propagation environment~\cite{C:Relay-Based_Blockage_and_Antenna_Misalignment_Mitigation_in_THz_Wireless_Communications,C:mmWave_Human_Blockage_at_73_GHz_with_a_Simple_Double_Knife_Edge_Diffraction_Model_and_Extension_for_Directional_Antennas}, reducing the ability of the THz signals to diffract around obstacles. As a result, people and objects that intercede the THz signal propagation path can act as blockers, leading to significant signal attenuation~\cite{Mag:RIS_pathl_alex,C:Frequency_domain_scattering_loss_in_THz_band,Mag:Beamf_Eff_RIS_giorgos_2021,C:RIS_eff_giorgos_2022,C:Impact_RIS_size_beam_eff_giorgos_2021}.

To mitigate the effects of pathloss and blockage in THz wireless communications, the use of reconfigurable-intelligent-surfaces (RIS) has been proposed as a promising enabler to reinstate the interrupted THz links~\cite{Mag:Beamf_Eff_RIS_giorgos_2021,C:EGW_Imp_Beam_Misal_RIS_assist_THz_Sys,A:Joint_Imp_Ph_err_HD_imp_mob_Int_RIS_over_kappa_mu,C:Eff_Rate_of_RIS_Netw_Loc_and_Ph_Est_Unc,C:OP_Anal_of_RIS_Wir_Netw_W_Von_Mises_Ph_Er,C:RIS_eff_giorgos_2022,C:Impact_RIS_size_beam_eff_giorgos_2021,A:Giorgos_Opt_pos_orient_RIS_mob_user,stratidakis2021analytical,BADARNEH2022,A:Alx_Casc_Comp_Turb_Misal_RIS,A:Ex_anal_RIS_aided_THz_over_a_m_point_errors,2020arXiv201200267D,A:Alx_OP_RIS_UAV_Misal,C:Alx_Perf_Anal_Multi_RIS_Surf_Emp_THz_Wir_Sys}. In~\cite{C:EGW_Imp_Beam_Misal_RIS_assist_THz_Sys}, a RIS assisted THz link is assumed, where the main lobe of the receiver beam is impaired by a stochastic Gaussian error. The performance is quantified in terms of average signal-to-noise-ratio (SNR), ergodic capacity and outage probability (OP). In~\cite{A:Joint_Imp_Ph_err_HD_imp_mob_Int_RIS_over_kappa_mu}, a RIS aided wireless system is considered where the destination node is surrounded by mobile interferers. The channels of the individual links of the system components with the RIS are modeled by the ${\kappa\text{--}\mu}$ distribution and the transceivers hardware is assumed to be non ideal. The system performance was evaluated in terms of coverage probability, ergodic capacity and signal to interference plus distortion ratio. In~\cite{C:Eff_Rate_of_RIS_Netw_Loc_and_Ph_Est_Unc}, a ring shaped RIS assisted wireless network was assumed. The transmitter was equipped with a low cost and accuracy hardware, which led to very poor phase estimation accuracy. The system performance was examined by means of the effective rate. In~\cite{C:OP_Anal_of_RIS_Wir_Netw_W_Von_Mises_Ph_Er}, the OP performance of a RIS assisted wireless system was investigated. The RIS elements were assumed to be imperfect and the resulting phase error of each element was modeled by a Von Mises distribution. In~\cite{C:RIS_eff_giorgos_2022}, the efficiency of a RIS assisted THz link is quantified in terms of the received power. This study was focused into the RIS orientation in the system and the transmitter beam was modeled as a Gaussian beam. Both the cases where the RIS was under full and partial illumination by the transmitted beam were investigated. It was observed that for both RIS illumination scenarios the RIS orientation relative to receiver position significantly impacted the system performance. Also, the system performance was affected by the transmitter position under the different RIS illumination regimes. Meanwhile,~\cite{Mag:Beamf_Eff_RIS_giorgos_2021,C:Impact_RIS_size_beam_eff_giorgos_2021} provided a generalized approach for the beamforming efficiency in terms of received power for a RIS assisted THz wireless system. In more detail, the beam of the transmitter was assumed to be Gaussian and the system performance was qualitatively divided into two regimes of operation, regarding the RIS size relative to the footprint of the transmitter beam on the RIS surface. Then, for a small footprint a tractable analytical model was derived. This model showcased the interplay among crucial system parameters, such as the positioning of the RIS with respect to the transmitter and the receiver and the properties of the transited beam. By considering a large beam footprint on the RIS surface the effect of the RIS size on the maximum received power was demonstrated by means of simulations. In~\cite{A:Giorgos_Opt_pos_orient_RIS_mob_user,stratidakis2021analytical}, the authors evaluated the different received power levels available for a RIS assisted THz wireless link, where the receiver was mobile. To determine the RIS position that maximized the received power the interplay of the crucial parameters of the available transceiver gains and a minimum requirement of received power were taken into account. In~\cite{BADARNEH2022}, a RIS assisted THz wireless system was assumed, where the channel followed the ${\kappa\text{--}\mu}$ shadowed fading model. The impact of the channel parameters, the number of RIS reflecting elements and the misalignment in the form of pointg error was performed by means of OP, ergodic capacity and average BER. Meanwhile, the authors in~\cite{A:Alx_Casc_Comp_Turb_Misal_RIS} introduced the theoretical framework for statistically characterizing the cascaded composite atmospheric turbulence and beam misalignment in the form of pointing errors, for both the scenarios of multiple RIS empowered free-space-optics and THz wireless systems. Then, by extracting the channel distributions of the two scenarios, their performance was evaluated by means of the OP. In~\cite{A:Ex_anal_RIS_aided_THz_over_a_m_point_errors}, a closed form expression for the SNR of a RIS aided THz wireless system was derived. In particular the transmitter-RIS and RIS-receiver links experienced ${\alpha\text{--}\mu}$ fading, the RIS-receiver link was impaired by beam misalignment in the form of pointing error, while non-ideal transceiver hardware was assumed. The system performance was evaluated by means of OP and bit-error-rate (BER) closed form expressions. In~\cite{2020arXiv201200267D}, a method for the optimization of the phase shifts of the RIS elements is proposed. Then, the statistical characterization of the SNR of a RIS aided THz wireless system was derived and the performance was evaluated by means of OP and ergodic capacity analytical expressions. In~\cite{A:Alx_OP_RIS_UAV_Misal}, the performance of a RIS assisted THz wireless unmanned-aerial-vehicle (UAV) system was analytically evaluated in terms of OP. The transmitter-RIS and RIS-UAV wireless channels are modeled by means of independent mixture-Gamma distributions. Also the OP performance was evaluated for ideal RIS-UAV channel and the case where it experienced not only mixture-Gamma small-scale fading, but also disorientation, beam misalignment in the form of pointing error, and transceivers hardware impairments. In~\cite{C:Alx_Perf_Anal_Multi_RIS_Surf_Emp_THz_Wir_Sys}, a theoretical framework for the performance analysis in terms of OP for a THz wireless system aided by multiple RISs, which was impaired by atmospheric turbulence, beam misalignment in the form of pointing error and transceivers hardware impairments was introduced.

\subsection{Motivation and Contribution}\label{Sec:Motivation}
In view of the rapidly evolving technologies in wireless networks, utilization of the THz band promises to provide the necessary high bandwidth. At such high frequency bands, directive beams for mitigating the pathloss and RISs for restoring the line-of-sight become an integral part, however their combined use is still largely unexplored. Therefore, it is of central importance to understand the stochastic behavior of misalignment between the RIS and the user and to obtain analytical models that can clarify the crucial bounds imposed by the system design parameters on the link performance. In this work the impact of beam misalignment in RIS-aided THz links is studied theoretically and the necessary analytical models for the stochastic characterization of the THz channel are derived. The main contributions are summarized as follows.
\begin{itemize}
  \item The system performance is studied into two distinct regimes, namely misalignment that occurs (a) on the steering plane and (b) normally to the steering plane.
  \item Analytical models are derived for both regimes, the validity of which is verified through numerical calculations.
  \item The impact of misalignment on the link robustness is explained by means of the analytical models for both regimes.
  \item The stochastic RIS performance is assessed with respect to crucial link parameters, such as the transmitter's beam width, the transmitter-RIS distance, the RIS-receiver distance, and the steering angle of the RIS.
\end{itemize}
\subsection{Organization and notations}
The remainder of this paper is organized as follows. Section~\ref{sec:SM} presents the RIS assisted THz wireless link system model. In Section~\ref{Sec:SNR_at_UE} the SNR is determined at the position of the UE, i.e. in the absence of misalignment, which serves as the maximum available SNR when misalignment is later introduced. In Section~\ref{Sec:SNR_dist} misalignment between the RIS reflected beam and the user is introduced and is studied under stochastic errors. The analytical SNR distribution is derived for two misalignment regimes, namely when stochastic misalignment error occurs on the steering plane (Section~\ref{Sec:Case4}) or normally to the steering plane (Section~\ref{Sec:Case3}). Section~\ref{Sec:Case4_Results} provides analytical and numerical results by employing the SNR distribution of the RIS beam misalignment scenario of Section~\ref{Sec:Case4}. Section~\ref{Sec:Compare_Case4_with_Case3} offers a qualitative comparison between the two misalignment regimes. Finally, Section~\ref{Sec:Coclusion} provides the concluding remarks of this work.
\\\\
Notations: In this paper the operators ${\mathbb{E}\left[\cdot\right]}$, ${\mathbb{V}\left[\cdot\right]}$, $\left|\cdot\right|$ stand for the statistical expectation, variance, and absolute value, respectively. The ${\sin{\left(\cdot\right)}}$, ${\cos{\left(\cdot\right)}}$, ${\sec{\left(\cdot\right)}}$ represent the trigonometric functions of sine, cosine and secant, respectively. Additionally, the mathematical operators ${\exp\left(\cdot\right)}$, ${\ln{\left(\cdot\right)}}$ and $\sqrt{\cdot}$ denote the exponential function, the natural logarithm and the square root. Finally, $\erf{\left(\cdot\right)}$ is the error function~\cite[Eq. 4.93]{papoulis}, whereas ${N\left(\mu,\sigma^2\right)}$ stands for the Gaussian distribution with mean $\mu$ and standard deviation $\sigma$.
\section{System Model}\label{sec:SM}
\begin{figure*}[t]
\begin{align}
S_r=\frac{\frac{2 P_t \left|R\right|^2}{\pi w_{\mathrm{RIS}}^2}}{\sqrt{\left(1+\frac{z_{\mathrm{B}}^2}{z_R^2}\right)\left(1+\frac{z_{\mathrm{B}}^2}{z_R^2 \cos^4\left(\theta_{\mathrm{B}}\right)}\right)}} \exp\left[-\frac{ko}{z_R}\left(\frac{x_{\mathrm{B}}^2+y_{\mathrm{B}}^2}{1+\frac{z_{\mathrm{B}}^2}{z_R^2}}-\left(1-\cos^4\left(\theta_{\mathrm{B}}\right)\right)\frac{\left(x_{\mathrm{B}} \cos\left(\phi_{\mathrm{B}}\right)+y_{\mathrm{B}}\sin\left(\phi_{\mathrm{B}}\right)\right)^2}{\left(1+\frac{z_{\mathrm{B}}^2}{z_R^2}\right)\left(1+\frac{z_R^2\cos^4\left(\theta_{\mathrm{B}}\right)}{z_{\mathrm{B}}^2}\right)}\right)\right]
\label{Eq:Srinf}
\end{align}
\end{figure*}
\begin{figure}[t!]
    \centering
    \includegraphics[width=0.95\columnwidth,trim=0 0 0 0,clip=false]{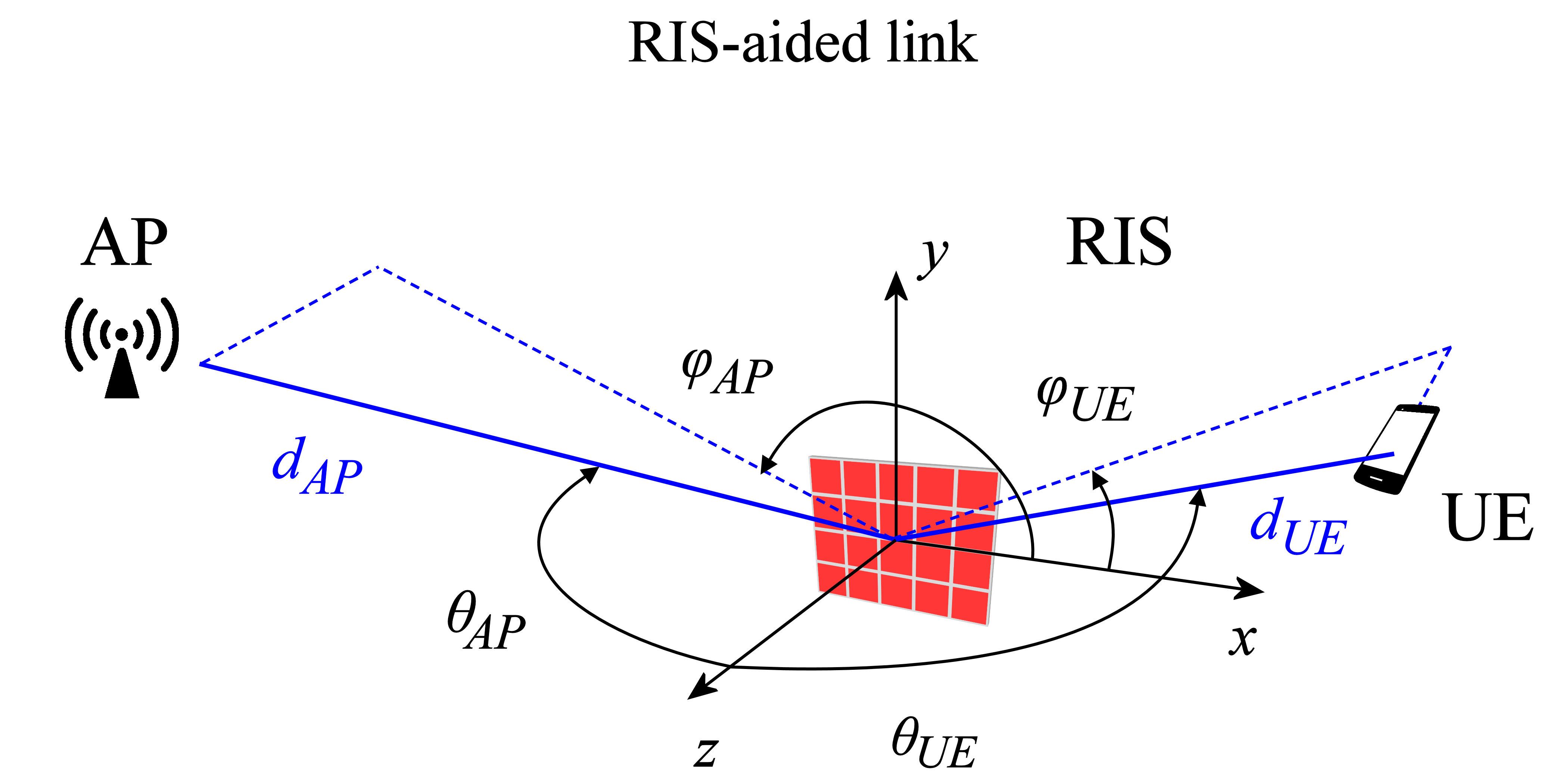}
    \caption{System model of RIS-aided link, illustrating the positions and relative angles between the AP, RIS and UE.}
    \label{fig:Sys_Model}
\end{figure}

%
%
%
A THz wireless system is considered, where the direct LoS link between the AP and the UE is blocked. To restore the AP--UE communication a RIS is employed, establishing LoS links with both the AP and the UE. The RIS is placed at the origin of the coordinate system shown in Fig.~\ref{fig:Sys_Model} and steers the incident beam towards any desired direction~\cite{Mag:Beamf_Eff_RIS_giorgos_2021,Mag:RIS_pathl_alex}. The AP is at distance $d_{\mathrm{AP}}$ from the RIS center and its position is characterized by the elevation and azimuth angles $\theta_{\mathrm{AP}}$ and ${\phi_{\mathrm{AP}}}$, respectively, defined on the global coordinate system ${xyz}$. The distance between the UE and the RIS center is $d_{\mathrm{UE}}$ and the corresponding elevation and azimuth angles are $\theta_{\mathrm{UE}}$ and ${\phi_{\mathrm{UE}}}$, respectively. Both the AP and UE are equipped with high-gain antennas characterized by gains $G_t$ and $G_r$, respectively. Following the approach previously adopted in ~\cite{Mag:Beamf_Eff_RIS_giorgos_2021}, the main lobe of the AP beam is modelled by a Gaussian beam of tunable width and the RIS is considered to be sufficiently large to capture the incident beam, without significantly truncating it at the finite RIS boundaries ~\cite{Mag:Beamf_Eff_RIS_giorgos_2021,C:RIS_eff_giorgos_2022,C:Impact_RIS_size_beam_eff_giorgos_2021,A:Giorgos_Opt_pos_orient_RIS_mob_user,stratidakis2021analytical} (this case is usually referred to as \textit{partial illumination}). As a result, the footprint of the AP beam on the RIS is also Gaussian with power density given by~\eqref{Eq:Srinf}. The parameters $P_t$, $w_{\mathrm{RIS}}$ and $R$ stand for the total transmitted power of the AP beam, the radius of the beam footprint on the RIS surface and the common reflection coefficient of all RIS elements, respectively. The Rayleigh length, $z_R$, is expressed as ~\cite{Mag:Beamf_Eff_RIS_giorgos_2021}:
\begin{align}
z_R=\frac{k_o w_{\mathrm{RIS}}^2}{2}=\frac{4 k_o d^2_{\mathrm{AP}}}{G_{\mathrm{AP}}},
\label{Eq:z_R}
\end{align} 
where ${k_o=2 \pi/\lambda}$ is the free-space wavenumber, $\lambda$ is the corresponding wavelength, and $G_{\mathrm{AP}}$ stands for the gain of the AP antenna. According to~\eqref{Eq:Srinf}, the AP beam footprint is reflected by the RIS and propagates as a tilted Gaussian beam along the direction $z_B$, which is characterized by the elevation and azimuth angles $\theta_{\mathrm{B}}$ and $\phi_{\mathrm{B}}$, respectively. The parameters $x_{\mathrm{B}}$, $y_{\mathrm{B}}$, $z_{\mathrm{B}}$ in~\eqref{Eq:Srinf} designate the non-orthogonal local beam coordinate system that follows the beam as it propagates distance $z_{\mathrm{B}}$ from the center of the RIS ($x_{\mathrm{B}}\parallel x$, $y_{\mathrm{B}}\parallel y$) and are expressed in terms of angles $\theta_{\mathrm{B}}, \phi_{\mathrm{B}}$ as:
\begin{align}
x_{\mathrm{B}}=x_0-z_{\mathrm{B}} \sin\left(\theta_{\mathrm{B}}\right)\cos\left(\phi_{\mathrm{B}}\right),
\label{Eq:x_B}
\end{align}
\begin{align}
y_{\mathrm{B}}=y_0-z_{\mathrm{B}}\sin\left(\theta_{\mathrm{B}}\right)\sin\left(\phi_{\mathrm{B}}\right),
\label{Eq:y_B}
\end{align} 
\begin{align}
z_{\mathrm{B}}=\frac{z_0}{\cos\left(\theta_{\mathrm{B}}\right)},
\label{Eq:z_B}
\end{align}
where
${\left(x_0,y_0,z_0\right)}$ are the Cartesian coordinates of the observation point at which $S_r$ is calculated. In spherical coordinates the location of the observation point can be expressed as $x_0= r_0\sin{\theta_0}\cos{\phi_0}, y_0=r_0\sin{\theta_0}\sin{\phi_0}, z_0=r_0\cos{\theta_0}$, where ${r_0=\sqrt{x_0^2+y_0^2+z_0^2}}$, and $\theta_0,\phi_0$ are the elevation and azimuth angles with respect to the RIS center.
\section{SNR at the position of the UE}\label{Sec:SNR_at_UE}
%
%
%
%
%
%
%
The SNR at the observation point ${\left(x_0,y_0,z_0\right)}$, is expressed by means of~\eqref{Eq:Srinf} as
\begin{align}
\mathrm{SNR}=\frac{S_r A_r}{N_o},
\label{Eq:SNR_1}
\end{align}
where $N_o$ stands for the additive white Gaussian noise power density and ${A_r=\frac{G_r \lambda^2}{4 \pi}}$ is the aperture of the receiver antenna.

In the absence of misalignment the beam reflected by the RIS is directed exactly towards the UE, i.e. ${\theta_{\mathrm{B}}=\theta_{\mathrm{UE}}}$ and ${\phi_{\mathrm{B}}=\phi_{\mathrm{UE}}}$, which is located at ${\left(x_{\mathrm{UE}},y_{\mathrm{UE}},z_{\mathrm{UE}}\right)}$ with:
\begin{align}
x_{\mathrm{UE}}=d_{\mathrm{UE}}\sin\left(\theta_{\mathrm{UE}}\right)\cos\left(\phi_{\mathrm{UE}}\right),
\label{Eq:x_UE}
\end{align}
\begin{align}
y_{\mathrm{UE}}=d_{\mathrm{UE}}\sin\left(\theta_{\mathrm{UE}}\right)\sin\left(\phi_{\mathrm{UE}}\right),
\label{Eq:y_UE}
\end{align}
\begin{align}
z_{\mathrm{UE}}=d_{\mathrm{UE}}\cos\left(\theta_{\mathrm{UE}}\right).
\label{Eq:z_UE}
\end{align}
The SNR at the location of the UE can be obtained from~\eqref{Eq:SNR_1} with~${\eqref{Eq:x_B}\text{--}\eqref{Eq:z_B}}$ calculated for an observation point exactly at the UE position, i.e. at ${x_0=x_{\mathrm{UE}}}$, ${y_0=y_{\mathrm{UE}}}$, and ${z_0=z_{\mathrm{UE}}}$. Using ${\eqref{Eq:x_UE}\text{--}\eqref{Eq:z_UE}}$ leads to ${x_{\mathrm{B}}=0}$, ${y_{\mathrm{B}}=0}$, and ${z_{\mathrm{B}}=d_{\mathrm{UE}}}$ and the SNR at the position of the UE is given by:
\begin{align}
\mathrm{SNR}=\frac{2 P_t \left|R\right|^2 A_r}
{N_o \pi w_{\mathrm{RIS}}^2
\sqrt{\left(1+\frac{d_{\mathrm{UE}}^2}{z_R^2}\right)
+\left(1+\frac{d_{\mathrm{UE}}^2}{z_R^2 \cos^4{\left(\theta_{\mathrm{UE}}\right)}}\right)
}}.
\label{Eq:SNR_at_UE}
\end{align}
In essence, exactly at the UE position, the argument of the exponential term in~\eqref{Eq:Srinf} becomes zero, leading to a maximum power density expressed by the remaining prefactor. For any other beam direction, the exponential term in~\eqref{Eq:Srinf} is always smaller than unity, leading to reduced SNR and, therefore, the SNR obtained from~\eqref{Eq:SNR_at_UE} stands for the maximum SNR at the UE in the absence of misalignment.

\section{SNR distribution in the presence of misalignment}\label{Sec:SNR_dist}
%
%
%
\begin{figure}[t!]
    \centering
    \includegraphics[width=0.95\columnwidth,trim=0 0 0 0,clip=false]{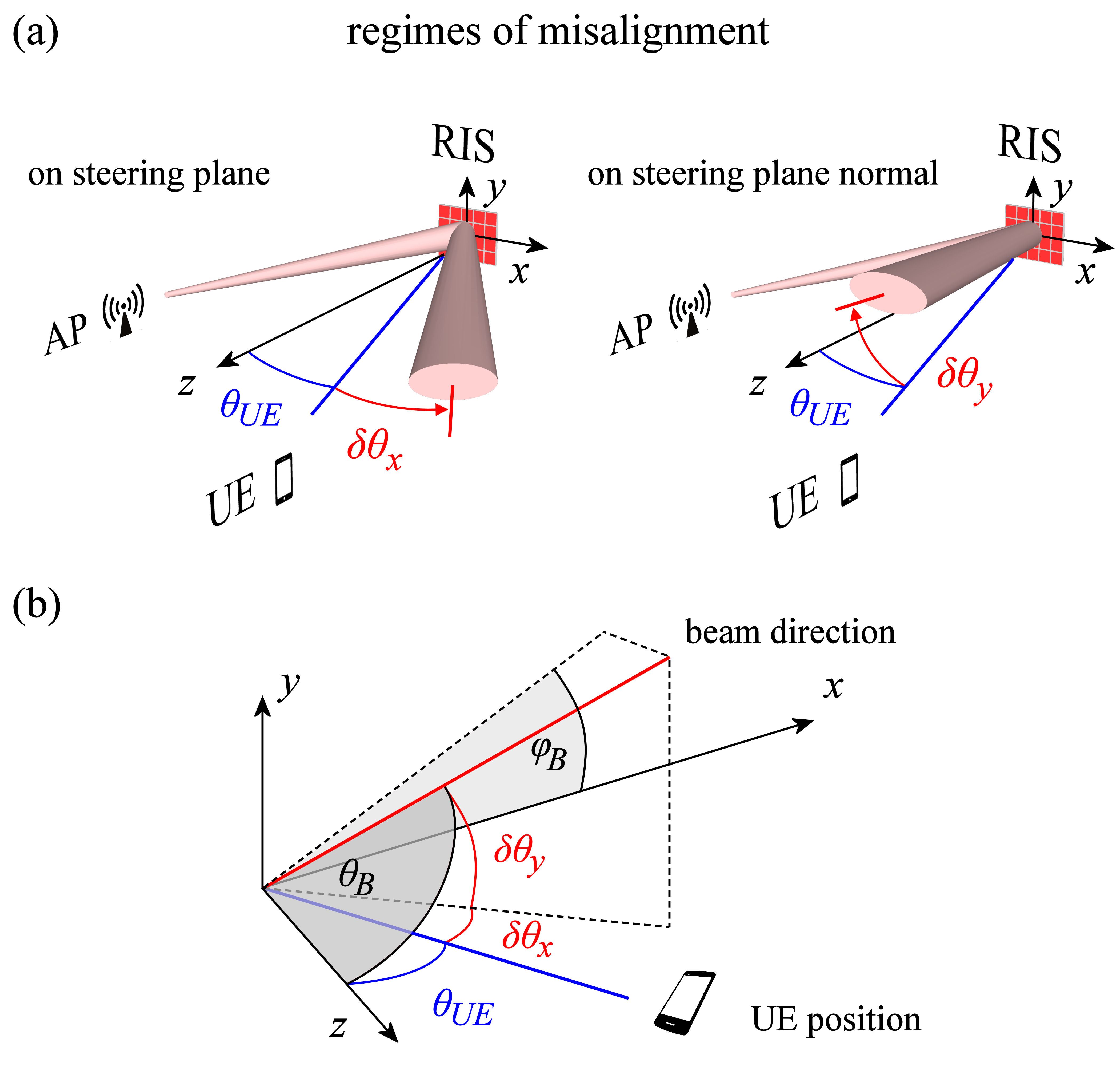}
    \caption{Misalignment between the beam reflected by the RIS and the UE. (a) Schematic illustration of misalignment on the $xz$-plane (left) and on the normal plane passing through the $y$-axis (right). The UE is located along the solid blue line, which is characterized by the angles $\theta_{\mathrm{UE}}$, $\phi_{\mathrm{UE}}$ and the beam is directed along the solid red line, characterized by the angles $\theta_{\mathrm{B}}$, $\phi_{\mathrm{B}}$. The reflected beam acquires elliptical shape, with distinct implications on the AP-UE link robustness for each regime. (b) Detailed depiction of error angles $\delta\theta_x, \delta\theta_y$, associated with the misalignment between the beam direction and the UE position. For $\delta\theta_x=\delta\theta_y=0$, misalignment is suppressed and the RIS steers the beam towards the UE with ${\theta_{\mathrm{B}}=\theta_{\mathrm{UE}}}$.}
    \label{fig:Sys_Model_misal}
\end{figure}
%
%
The case where the RIS reflected beam does not accurately point in the direction of the UE signifies the phenomenon of misalignment. This could result from fast UE motion that leads to poor beam tracking or even by the RIS operation itself, i.e. fabrication imperfections or limited range of steering angles due to the particular design (e.g. access to quantized phase shifts that limit the RIS performance~\cite{C:EGW_Imp_Beam_Misal_RIS_assist_THz_Sys,A:Joint_Imp_Ph_err_HD_imp_mob_Int_RIS_over_kappa_mu,A:RIS_Disc_Ph_Shift,A:Beam_Opt_RIS_Disc_Ph_Shits}). In case where the misalignment is predictable,~\eqref{Eq:SNR_1} provides the expected SNR at the UE. What is more interesting for wireless links, though, is when the RIS misalignment has stochastic behavior and, to study the RIS-aided link in this case, it is necessary to derive the probability density function (PDF) and the cumulative density function (CDF) of the SNR. Without loss of generality, and as also considered in previous works~\cite{Mag:Beamf_Eff_RIS_giorgos_2021,C:RIS_eff_giorgos_2022,C:Impact_RIS_size_beam_eff_giorgos_2021,A:Giorgos_Opt_pos_orient_RIS_mob_user,stratidakis2021analytical}, the UE is located on the ${x\text{--}z}$ plane, where steering takes place. In this work, misalignment is studied for errors occurring either on the steering plane or on the steering plane normal that passes through the $y$-axis, as shown in Fig.~\ref{fig:Sys_Model_misal}(a). Moreover, as shown in Fig.~\ref{fig:Sys_Model_misal}(b), these errors are expressed in terms of angles ${\delta\theta_x \sim N\left(0,\sigma_{\theta_x}^2\right)}$ and ${\delta\theta_y \sim N\left(0,\sigma_{\theta_y}^2\right)}$, respectively, where $\sigma_{\theta_x}$, $\sigma_{\theta_y}$ stand for the standard deviation in each case. In general, while misalignment errors may span from a few degrees up to a wide angular range, what is here more relevant is small deviations from the desired steering angle. Therefore, it is reasonable to study misalignment for relatively small errors and the analytical PDF and CDF expressions for the SNR presented in this work are derived by performing asymptotic analysis on~\eqref{Eq:SNR_1}.
\subsection{Misalignment on the steering plane}\label{Sec:Case4}

For a UE located on the ${x\text{--}z}$ plane, i.e. ${\phi_{\mathrm{UE}}=0^o}$, the beam is steered towards ${\theta_{\mathrm{B}}=\theta_{\mathrm{UE}}}$, ${\phi_{\mathrm{B}}=0^o}$. For misalignment occurring on the same plane (${\delta\theta_x \neq 0^o}$ and ${\delta\theta_y=0^o}$) the angle $\delta\theta_x$ corresponds to errors on the elevation angle $\theta_{\mathrm{B}}$, i.e. the steering angle can be written as ${\theta_{\mathrm{B}}=\theta_{\mathrm{UE}}+\delta\theta_x}$. Then, the approximation of~\eqref{Eq:SNR_1} can be obtained~as
\begin{align}
\widetilde{\mathrm{SNR}}=\alpha \exp\left(-\beta \delta\theta_x^2\right),
\label{Eq:SNR_form_c1_c3_c4}
\end{align}
where
\begin{align}
\alpha=\frac{2 P_t \left|R\right|^2 z_R^2 A_r}
{N_o \pi w_{\mathrm{RIS}}^2 \sqrt{\left(d_{\mathrm{UE}}^2+z_R^2\right)\left(z_R^2+\frac{d_{\mathrm{UE}}^2}{\cos^4\left(\theta_{\mathrm{UE}}\right)}\right)}}
\label{Eq:a_4}
\end{align}
and
\begin{align}
\beta=\frac{\frac{d_{\mathrm{UE}}^2k_oz_R}{\cos^2\left(\theta_{\mathrm{UE}}\right)}}{z_R^2+\frac{d_{\mathrm{UE}}^2}{\cos^4\left(\theta_{\mathrm{UE}}\right)}}.
\label{Eq:b_4}
\end{align}
In the following~\eqref{Eq:f_SNR_c1_c3_c4_s2}--\eqref{Eq:Skewness_Case4} the PDF, CDF, average SNR, variance and skewness of $\widetilde{\mathrm{SNR}}$ are obtained as a function of the $\alpha$ and $\beta$. In more detail, the PDF corresponding to~\eqref{Eq:SNR_form_c1_c3_c4} can be obtained~as
\begin{align}
f_{\widetilde{\mathrm{SNR}}}\left(x\right)=\frac{\left(\frac{x}{\alpha}\right)^{\frac{1}{2 \beta \sigma_{\theta_{\mathrm{B}}}^2}}}{\sqrt{2\pi}\sigma_{\theta_{\mathrm{B}}} x \beta\sqrt{\frac{\left|\ln\left(\frac{x}{\alpha}\right)\right|}{\beta}}}
\label{Eq:f_SNR_c1_c3_c4_s2}
\end{align}
and the CDF can be obtained~as
\begin{align}
F_{\widetilde{\mathrm{SNR}}}\left(x\right)=1-\erf\left(\frac{\sqrt{\ln\left(\frac{\alpha}{x}\right)}}{\sqrt{2\beta}\sigma_{\theta_{\mathrm{B}}}}\right),
\label{Eq:F_SNR_c1_c3_c4_s2}
\end{align}
where $x$ is an instance of ${\widetilde{\mathrm{SNR}}}$ and ${x\in\left(0,\alpha\right)}$. The mean of ${\widetilde{\mathrm{SNR}}}$ can be obtained~as
\begin{align}
\overline{\mathrm{SNR}}=\frac{\alpha}{\sqrt{1+2\beta\sigma_{\theta_\mathrm{B}}^2}}.
\label{Eq:Mean_c1_c3_c4_s2}
\end{align}
The variance of $\widetilde{\mathrm{SNR}}$ can be obtained~as
\begin{align}
\sigma_{\widetilde{\mathrm{SNR}}}^2=\alpha^2
\left(\frac{1}{\sqrt{1+4\beta\sigma_{\theta_{\mathrm{B}}}^2}}+\frac{1}{-1-2\beta\sigma_{\theta_{\mathrm{B}}}^2}\right).
\label{Eq:Var_c1_c3_c4_s3}
\end{align}
Meanwhile, the skewness can be obtained~as
\begin{align}
\begin{split}
s_{\widetilde{\mathrm{SNR}}}&=\frac{
\frac{1}{\sqrt{1+6 \beta \sigma_{\theta_{\mathrm{B}}}^2}}
+
\frac{-3-6 \beta \sigma_{\theta_{\mathrm{B}}}^2+2 \sqrt{1+4 \beta \sigma_{\theta_{\mathrm{B}}}^2}}{\sqrt{1+4 \beta \sigma_{\theta_{\mathrm{B}}}^2} \left(1+2 \beta \sigma_{\theta_{\mathrm{B}}}^2\right)^{3/2}}
}
{
\left(\frac{1}{\sqrt{1+4 \beta \sigma_{\theta_{\mathrm{B}}}^2}}
+\frac{1}{-1-2 \beta \sigma_{\theta_{\mathrm{B}}}^2}\right)^{3/2}
}
\end{split}
\label{Eq:Skewness_Case4}
\end{align}
The proof of the derivation of~\eqref{Eq:SNR_form_c1_c3_c4}--\eqref{Eq:Skewness_Case4} can be found in Appendix A~\ref{Sec:Appendix_A}.

\subsection{Misalignment on steering plane normal}\label{Sec:Case3}

For misalignment occurring on the steering plane normal, $\delta\theta_x=0^o$ and ${\delta\theta_y \neq 0^o}$. In this case the angle $\delta\theta_y$ is expressed in terms of both $\theta_{\mathrm{B}}$ and $\phi_{\mathrm{B}}$ and the approximation of~\eqref{Eq:SNR_1} can be obtained~as
\begin{align}
\widetilde{\mathrm{SNR}}=\alpha
\exp\left(-\zeta \delta\theta_y^2
\right),
\label{Eq:SNR_case3_approx_s2}
\end{align}
where
\begin{align}
\zeta=\frac{k_o z_R d_{\mathrm{UE}}^2}{z_R^2+d_{\mathrm{UE}}^2}.
\label{Eq:b_3}
\end{align}
The parameter $\alpha$ is the same as in the previous case and is given by~\eqref{Eq:a_4}. The PDF, CDF, average SNR, variance and skewness of $\widetilde{\mathrm{SNR}}$ in this scenario can be obtained as a function of $\alpha$ and $\zeta$ as in~\eqref{Eq:f_SNR_c1_c3_c4_s2}--\eqref{Eq:Skewness_Case4}, where $\beta$~\eqref{Eq:b_4} is replaced by $\zeta$. The proof of the derivation of~\eqref{Eq:SNR_case3_approx_s2} and the aforementioned analytical expressions in this scenario can be found in Appendix B~\ref{Sec:Appendix_B}.
\section{Results}\label{Sec:Results}
\begin{figure}
\centering
\includegraphics[width=1\columnwidth]{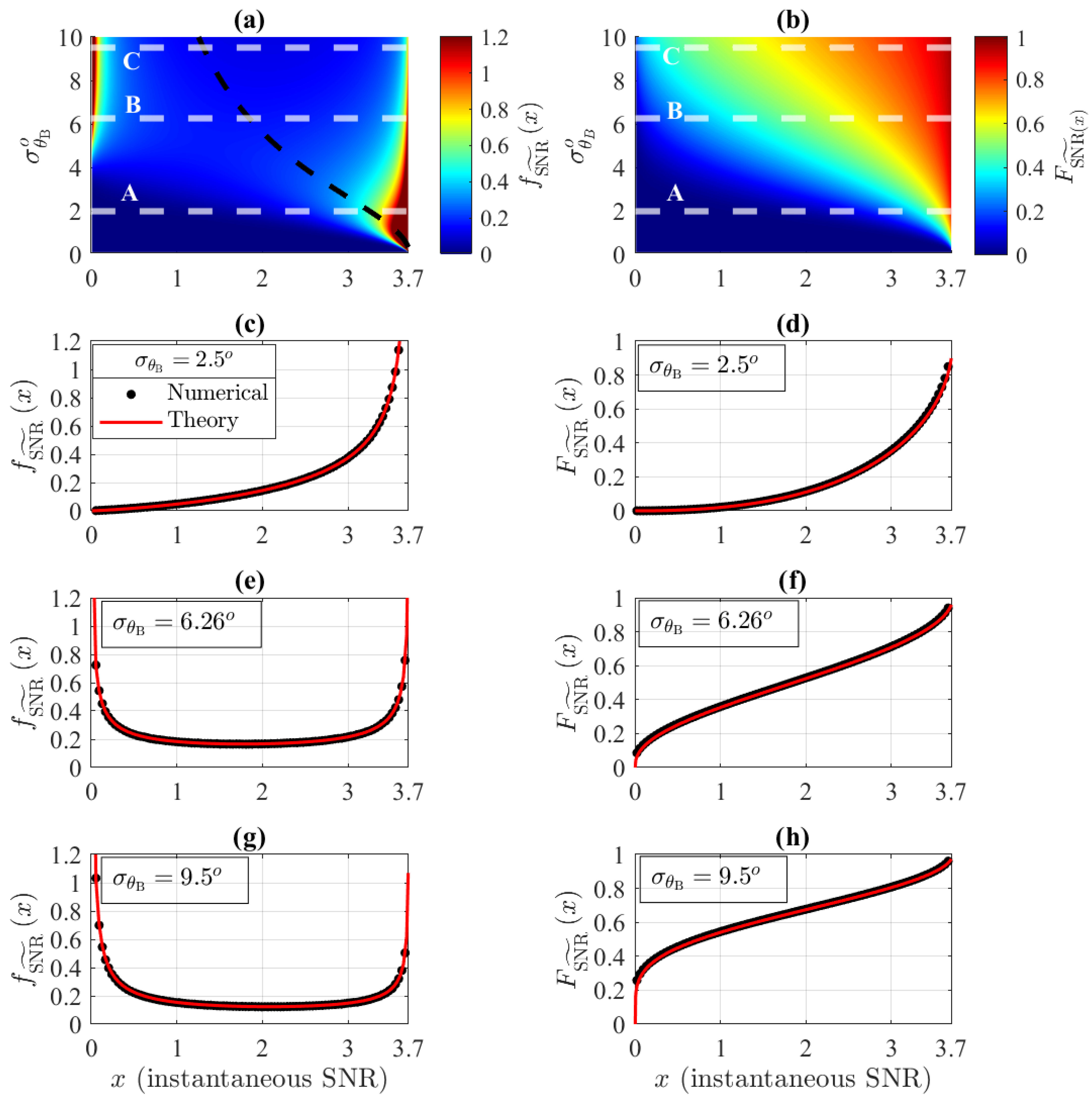}
\caption{(a),(b) ${f_{\widetilde{\mathrm{SNR}}}\left(x\right)}$~\eqref{Eq:f_SNR_c1_c3_c4_s2} and ${F_{\widetilde{\mathrm{SNR}}}\left(x\right)}$~\eqref{Eq:F_SNR_c1_c3_c4_s2} as function of $x$ and $\sigma_{\theta_{\mathrm{B}}}$, respectively. (c),(e),(g) ${f_{\widetilde{\mathrm{SNR}}}\left(x\right)}$ and numerical SNR PDF~\eqref{Eq:SNR_1} as a function of $x$, respectively. (d),(f),(h) ${F_{\widetilde{\mathrm{SNR}}}\left(x\right)}$ and numerical SNR CDF as a function of $x$, respectively. The AP beam footprint radius is ${w_{\mathrm{RIS}}=25\text{ cm}}$, and the UE is at distance ${d_{\mathrm{UE}}=2\text{ m}}$, with ${\theta_{\mathrm{UE}}=0^o}$.}
\label{Fig:PDF_CDF_figs}
\end{figure}
%
%
%
\subsection{Misalignment on the steering plane}\label{Sec:Case4_Results}
In this section numerical and theoretical results associated with the $\widetilde{\mathrm{SNR}}$ approximation distribution of~\eqref{Eq:SNR_form_c1_c3_c4} in Section~\ref{Sec:Case4} are presented. In more detail, Figs.~\ref{Fig:PDF_CDF_figs}--\ref{fig:Fig_5} show PDF, CDF, $\overline{\mathrm{SNR}}$, and skewness results for different combinations of values of the $w_{\mathrm{RIS}}$, $d_{\mathrm{UE}}$, $\sigma_{\theta_{\mathrm{B}}}$, and $\theta_{\mathrm{UE}}$ parameters of the system and different levels of misalignment. These results are obtained from the analytical expressions ${\mathrm{~\eqref{Eq:f_SNR_c1_c3_c4_s2}\text{--}\eqref{Eq:Mean_c1_c3_c4_s2}}}$ and~\eqref{Eq:Skewness_Case4} and have been also validated with numerical evaluation of the full SNR expression~\eqref{Eq:SNR_1} using ${N=10^7}$ distribution samples for each operation point. It should be noted that in the following figures unless otherwise is stated, the continuous red lines indicate the analytically obtained PDF and CDF expressions according to~\eqref{Eq:f_SNR_c1_c3_c4_s2} and~\eqref{Eq:F_SNR_c1_c3_c4_s2}, respectively, whereas the black dots stand for the numerically obtained PDF and CDF of the total SNR expression of~\eqref{Eq:SNR_1}. Also, unless otherwise is stated for the following results it was assumed that ${\theta_{\mathrm{UE}}=0^o}$, ${w_{\mathrm{RIS}}=25\text{ cm}}$, ${d_{\mathrm{UE}}=2\text{ m}}$, ${G_r=40\text{ dB}}$, ${f=140\text{ GHz}}$, ${\frac{P_t}{N_o}=20\text{ dB}}$, ${\left|R\right|=1}$.
\subsubsection{PDF and CDF of SNR}
Fig.~\ref{Fig:PDF_CDF_figs}(a), illustrates ${f_{\widetilde{\mathrm{SNR}}}\left(x\right)}$~\eqref{Eq:f_SNR_c1_c3_c4_s2} as a function of the instantaneous SNR $x$ and $\sigma_{\theta_{\mathrm{B}}}$. The colormap of ${f_{\widetilde{\mathrm{SNR}}}\left(x\right)}$ has been saturated within the range ${0\text{--}1.2}$ in order to emphasize the qualitative features. The dashed black line represents ${\overline{\mathrm{SNR}}}$ as a function of $\sigma_{\theta_{\mathrm{B}}}$. The white dashed lines A, B and C denote the cases where $\sigma_{\theta_{\mathrm{B}}} = $ ${2.5^o}$, ${6.26^o}$ and ${9.5^o}$, respectively. These aforementioned values of $\sigma_{\theta_{\mathrm{B}}}$, indicate characteristic levels of RIS reflected beam misalignment severity and give insights about the non-zero amplitudes of ${f_{\mathrm{\widetilde{SNR}}}\left(x\right)}$ in regards with the corresponding range of the instantaneous SNR. In this direction, the Figs.~\ref{Fig:PDF_CDF_figs}(c), (e), and (g) (corresponding to the dashed lines A, B and C of Fig.~\ref{Fig:PDF_CDF_figs}(a)) serve as an illustrative example, where ${f_{\widetilde{\mathrm{SNR}}}\left(x\right)}$ and the numerically obtained SNR PDF of~\eqref{Eq:SNR_1} are obtained as a function of the instantaneous SNR $x$. Moreover, Figs.~\ref{Fig:PDF_CDF_figs}(c), (e), and (g) motivate that, the analytical $f_{\widetilde{\mathrm{SNR}}}\left(x\right)$ expression can accurately approximate the numerical PDF of the full SNR expression. According to Fig.~\ref{Fig:PDF_CDF_figs}(a) for a low level of misalignment severity the non-zero amplitudes of ${f_{\widetilde{\mathrm{SNR}}}\left(x\right)}$ are obtained for high values of the instantaneous SNR. Fig.~\ref{Fig:PDF_CDF_figs}(c) is an indicative example of this observation. Furthermore, from Fig.~\ref{Fig:PDF_CDF_figs}(a) it is shown that by increasing $\sigma_{\theta_{\mathrm{B}}}$ up to a certain level, the non-zero values of ${f_{\widetilde{\mathrm{SNR}}}\left(x\right)}$ are located on both the low and high values of the instantaneous SNR. Fig.~\ref{Fig:PDF_CDF_figs}(e) is an indicative example of this remark. Moreover, as Fig.~\ref{Fig:PDF_CDF_figs}(a) indicates, by further increasing the misalignment severity i.e., $\sigma_{\theta_{\mathrm{B}}}$, the non-zero amplitudes of ${f_{\widetilde{\mathrm{SNR}}}\left(x\right)}$ are obtained for low values of the instantaneous SNR $x$. This observation is illustrated by Fig.~\ref{Fig:PDF_CDF_figs}(g).

Fig.~\ref{Fig:PDF_CDF_figs}(b), presents ${F_{\widetilde{\mathrm{SNR}}}\left(x\right)}$~\eqref{Eq:F_SNR_c1_c3_c4_s2} as a function of the instantaneous SNR, $x$ and $\sigma_{\theta_{\mathrm{B}}}$, and the white dashed lines A, B and C mark the corresponding cases studied in Fig.~\ref{Fig:PDF_CDF_figs}(a). It is observed that, for a given value of $x$, as $\sigma_{\theta_{\mathrm{B}}}$ increases ${F_{\widetilde{\mathrm{SNR}}}\left(x\right)}$ drastically increases. As an example, for ${x=2}$ by changing $\sigma_{\theta_{\mathrm{B}}}$ from $0.1^o$ to $5^o$ and $9^o$ the resulting ${F_{\widetilde{\mathrm{SNR}}}\left(x\right)}$ is $0$, $0.42$ and $0.66$, respectively. As a consequence the increase of the misalignment severity reduces the probability of the UE to operate above a predetermined SNR threshold. Furthermore, for a given $\sigma_{\theta_{\mathrm{B}}}$ and by increasing $x$, ${F_{\widetilde{\mathrm{SNR}}}\left(x\right)}$ drastically increases. For example, for ${\sigma_{\theta_{\mathrm{B}}}=3^o}$ and by setting $x$ equal to $0.5$, $3$ and $3.7$ the resulting ${F_{\widetilde{\mathrm{SNR}}}\left(x\right)}$ is $0.016$, $0.43$ and $0.9$, respectively. The aforementioned remarks made on the effect of $\sigma_{\theta_{\mathrm{B}}}$ and the instantaneous SNR on ${F_{\widetilde{\mathrm{SNR}}}\left(x\right)}$ can be further observed in Figs.~\ref{Fig:PDF_CDF_figs}(d), (f) and (h), which serve as illustrative examples. Figs.~\ref{Fig:PDF_CDF_figs}(d), (f) and (h) illustrate~\eqref{Eq:F_SNR_c1_c3_c4_s2} and the numerically obtained SNR CDF of~\eqref{Eq:SNR_1} as a function of the instantaneous SNR $x$, where ${\sigma_{\theta_{\mathrm{B}}}}$ was set equal to ${2.5^o}$, ${6.26^o}$ and ${9.5^o}$, respectively.
\subsubsection{Impact of AP beam footprint, RIS steering angle and RIS-UE distance on robustness of link on misalignment}
%
%
%
\begin{figure}
\centering
\includegraphics[width=0.9\columnwidth]{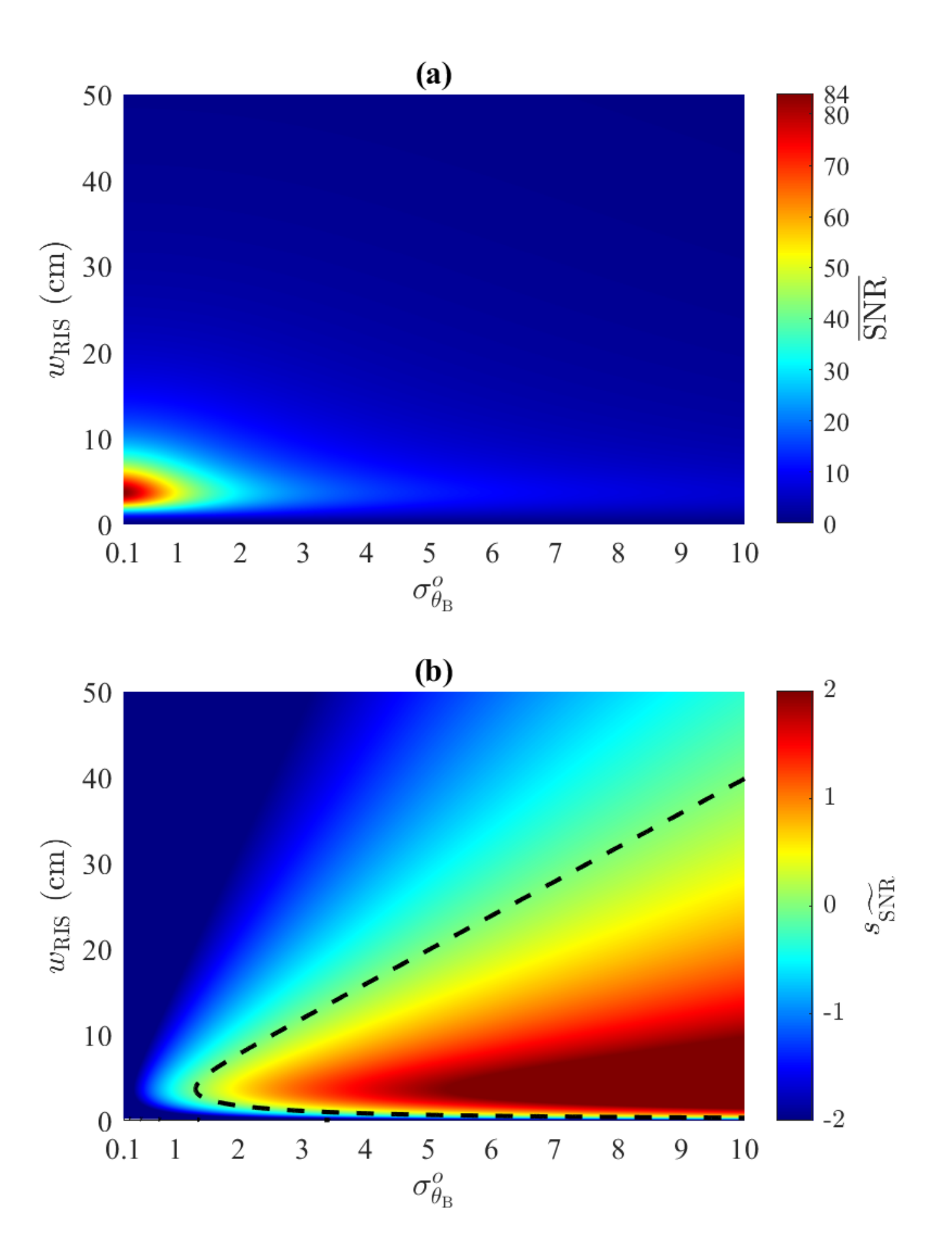}
\caption{Study of misalignment error for variable beam footprint, $w_{\mathrm{RIS}}$. (a) $\overline{\mathrm{SNR}}$ and (b) ${s_{\widetilde{\mathrm{SNR}}}}$, as a function of ${\sigma_{\theta_{\mathrm{B}}}}$. The user is at distance ${d_{\mathrm{UE}}=2\text{ m}}$ from the RIS, along the direction with ${\theta_{\mathrm{UE}}=0^o}$. The dashed line marks the locus of operation points with ${s_{\widetilde{\mathrm{SNR}}}}=0$.}
\label{fig:Fig_3}
\end{figure}
%
%
%
In the previous section the stochastic behavior of the RIS-aided link was studied for a certain configuration, i.e. for specific AP beam footprint, RIS steering angle and RIS-UE distance. These parameters play a crucial role on the robustness of the link on misalignment~\cite{Mag:Beamf_Eff_RIS_giorgos_2021} and, to gain further insight into the involved dependencies, in this section the stochastic behavior is studied as a function of these parameters. In the following examples the variance~\eqref{Eq:Var_c1_c3_c4_s3} of the ${\widetilde{\mathrm{SNR}}}$ approximation will be omitted, as it offers little insightful information on the SNR; notice how in Fig.~\ref{Fig:PDF_CDF_figs} the PDF changes qualitatively as the misalignment error increases. This change can be adequately described in terms of the skewness. In short, positive skewness indicates that the majority of the non-zero values of the analytical PDF expression~\eqref{Eq:f_SNR_c1_c3_c4_s2} are found on the higher range of the instantaneous SNR, whereas negative skewness shows that the majority of the non-zero values of~\eqref{Eq:f_SNR_c1_c3_c4_s2} are found on the lower range of the instantaneous SNR; with zero skewness, the non-zero values of~\eqref{Eq:f_SNR_c1_c3_c4_s2} tend to be equally spread between the low and high instantaneous SNR values.

Fig.~\ref{fig:Fig_3}(a), presents ${\overline{\mathrm{SNR}}}$~\eqref{Eq:Mean_c1_c3_c4_s2} as a function of $\sigma_{\theta_{\mathrm{B}}}$ and $w_{\mathrm{RIS}}$. The latter is determined by the gain of the transmitter's antenna and the transmitter-RIS distance, as expressed via~\eqref{Eq:z_R}. For typical antenna gains in the order of 20-30 dB and transmitter-RIS distances in the order of few meters, $w_{\mathrm{RIS}}$ is practically in the order of a few cm and, hence, the results are limited within the range ${w_{\mathrm{RIS}}\in\left(0,50\right]\text{ cm}}$. It is observed that, for any given $\sigma_{\theta_{\mathrm{B}}}$, as $w_{\mathrm{RIS}}$ increases the $\overline{\mathrm{SNR}}$  increases up to a certain footprint size, after which starts to decrease. According to~\cite{Mag:Beamf_Eff_RIS_giorgos_2021}, this is attributed to the transition from near to the far field of the reflected beam. In more detail, the maximum received power at the position of the UE is accomplished when the UE lies in the vicinity region of this transition~\cite{Mag:Beamf_Eff_RIS_giorgos_2021}. Meanwhile, for the greater values of ${w_{\mathrm{RIS}}}$ it is observed that, for a given $\sigma_{\theta_{\mathrm{B}}}$ and increasing $w_{\mathrm{RIS}}$ the $\overline{\mathrm{SNR}}$ is reduced. For example, for ${\sigma_{\theta_{\mathrm{B}}}=1^o}$ and changing $w_{\mathrm{RIS}}$ from $25$ to ${30 \text{ cm}}$ the resulting ${\overline{\mathrm{SNR}}}$ is $3.6$ and $2.5$, respectively. Also, for a given $w_{\mathrm{RIS}}$ as $\sigma_{\theta_{\mathrm{B}}}$, i.e. the misalignment severity increases and ${\overline{\mathrm{SNR}}}$ is reduced. Despite the aforementioned observations, a large beam footprint on the RIS mitigates the ${\overline{\mathrm{SNR}}}$ deterioration with the increase of $w_{\mathrm{RIS}}$. In more detail, for ${w_{\mathrm{RIS}}=23 \text{ cm}}$ and by changing $\sigma_{\theta_{\mathrm{B}}}$ from ${3^o}$ to ${4^o}$ and ${6^o}$ the corresponding $\overline{\mathrm{SNR}}$ is $3.25$, $2.82$ and $2.1$, respectively. Meanwhile, for ${w_{\mathrm{RIS}}=40 \text{ cm}}$ and by changing $\sigma_{\theta_{\mathrm{B}}}$ from ${3^o}$ to ${4^o}$ and ${6^o}$ the corresponding $\overline{\mathrm{SNR}}$ is $1.28$, $1.2$ and $1$, respectively. As a consequence, this misalignment robustness comes in the expense of the level of the available power capable of being detected by the UE and the corresponding $\overline{\mathrm{SNR}}$. Furthermore, as Fig.~\ref{fig:Fig_3}(a) illustrates, the higher ${\overline{\mathrm{SNR}}}$ is obtained for low values of $w_{\mathrm{RIS}}$ and $\sigma_{\theta_{\mathrm{B}}}$. For relatively wide footprints (larger than $w_{\mathrm{RIS}}\approx 5$ cm), the more the RIS impinging beam is focused the narrower the RIS reflected beam will be, which in turn increases the gain of reflected beam and hence $\overline{\mathrm{SNR}}$. As a consequence, a highly focused reflected beam combined with a low level of misalignment drastically improves the received power capable of being detected by the UE. For example, for ${w_{\mathrm{RIS}}=20\text{ cm}}$ and ${\sigma_{\theta_{\mathrm{B}}}=0.1^o}$ the resulting ${\overline{\mathrm{SNR}}=5.8}$. For relatively narrow footprints (smaller than $w_{\mathrm{RIS}}\approx 5$ cm), the situation is reversed. However, given, the currently achieved THz transceiver gains~\cite{Mag:High_Gain_D_Band_Transmitarrays,Mag:D_Band_H_Gain_Circ_Pol_Plate_Array_Ant,C:Two_Types_of_High_Gain_Slot_Array_Ant_based_on_Ridge_Gap_Waveguide_D_Band} and the expected transmission distances, from~\eqref{Eq:z_R} this situation is less expected in a practical RIS assisted THz wireless system. Hence, the results of ${\overline{\mathrm{SNR}}}$ obtained for especially low values of ${w_{\mathrm{RIS}}}$ are presented here only for the sake of completeness.

Fig.~\ref{fig:Fig_3}(b), illustrates ${s_{\widetilde{\mathrm{SNR}}}}$~\eqref{Eq:Skewness_Case4} as a function $\sigma_{\theta_\mathrm{B}}$ and $w_{\mathrm{RIS}}$. The dashed black line denotes the combination of ${\sigma_{\theta_{\mathrm{B}}}}$ and $w_{\mathrm{RIS}}$ values that yield ${s_{\widetilde{\mathrm{SNR}}}=0}$. For a given $w_{\mathrm{RIS}}$ as $\sigma_{\theta_{\mathrm{B}}}$ increases the skewness changes from negative to positive values. It is observed that, for a given $w_{\mathrm{RIS}}$ and low values of $\sigma_{\theta_{\mathrm{B}}}$ the skewness is negative. A negative skewness indicates that the majority of the non-zero values of ${f_{\widetilde{\mathrm{SNR}}}\left(x\right)}$~\eqref{Eq:f_SNR_c1_c3_c4_s2} are obtained for higher values of the instantaneous SNR. This becomes more prominent as the skewness values are further reduced. Fig.~\ref{Fig:PDF_CDF_figs}(c), serves as an illustrative example of this observation. Meanwhile, for a given $w_{\mathrm{RIS}}$ and intermediate values of $\sigma_{\theta_{\mathrm{B}}}$ the values of skewness are closer to zero both from the positive and negative direction. As the value of skewness tends to $0$ the non-zero amplitudes of~\eqref{Eq:f_SNR_c1_c3_c4_s2} spread almost evenly to the low and high instantaneous SNR. On the other hand, for the intermediate values of the instantaneous SNR the amplitude of~\eqref{Eq:f_SNR_c1_c3_c4_s2} is much lower compared to that obtained in the low and high instantaneous SNR regime. Fig.~\ref{Fig:PDF_CDF_figs}(e), serves as an illustrative example of this observation. For a given $w_{\mathrm{RIS}}$ and greater values of $\sigma_{\theta_{\mathrm{B}}}$ the skewness is positive. A positive skewness means that the majority of the non-zero amplitudes of~\eqref{Eq:f_SNR_c1_c3_c4_s2} are obtained for low values of the instantaneous SNR, which is more prominent with the increase of the skewness. Fig.~\ref{Fig:PDF_CDF_figs}(g), serves as an illustrative example of this observation.
%
%
%
\begin{figure}
\centering
\includegraphics[width=0.9\columnwidth]{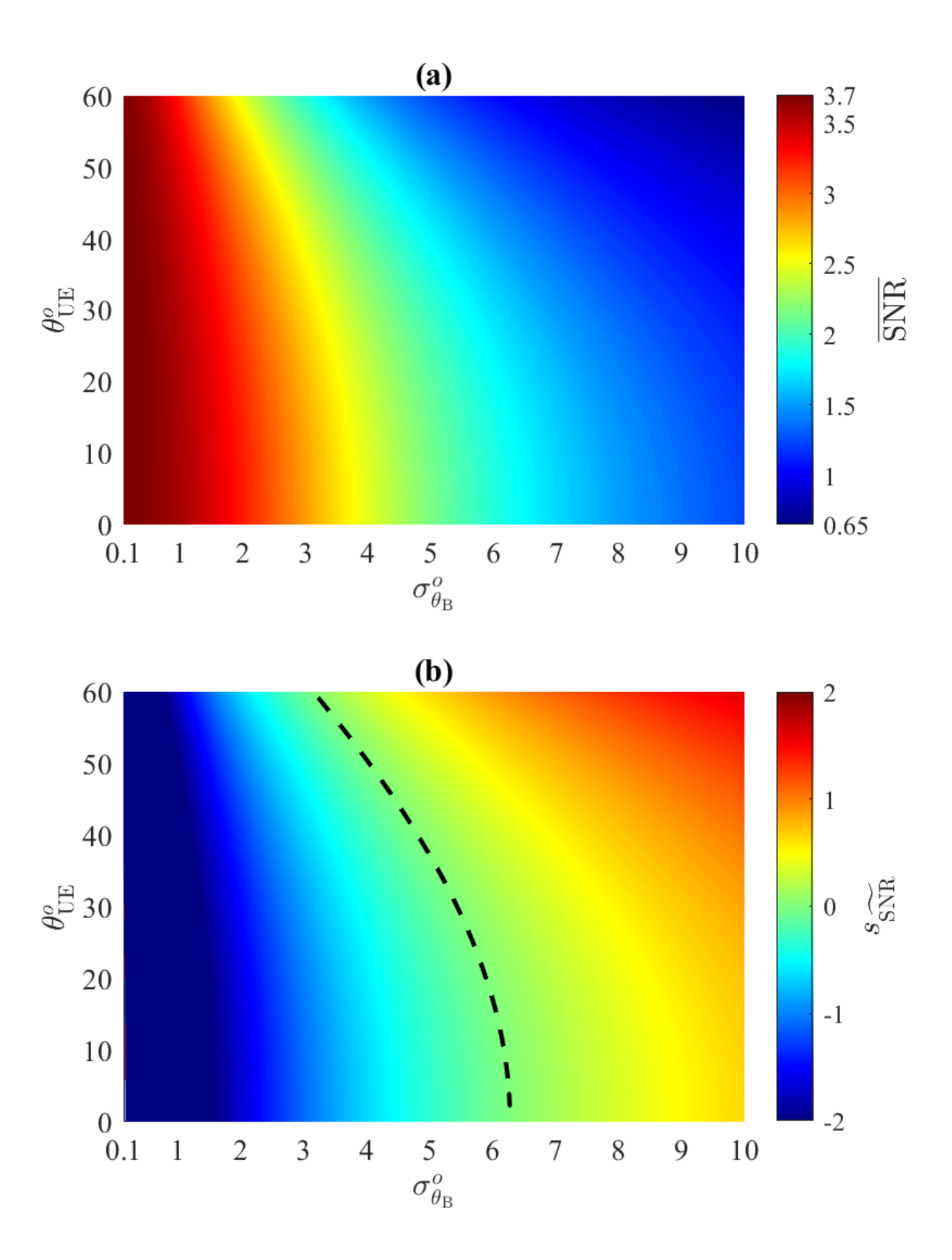}
\caption{Study of misalignment error for beam steering towards variable angle $\theta_{\mathrm{UE}}$. (a) $\overline{\mathrm{SNR}}$ and (b) ${s_{\widetilde{\mathrm{SNR}}}}$, as a function of ${\sigma_{\theta_{\mathrm{B}}}}$. The user is at distance ${d_{\mathrm{UE}}=2\text{ m}}$ from the RIS, and the footprint of the incident beam is ${w_{\mathrm{RIS}}=25\text{ cm}}$. The dashed line marks the locus of operation points with ${s_{\widetilde{\mathrm{SNR}}}}=0$.}
\label{fig:Fig_4}
\end{figure}

Fig.~\ref{fig:Fig_4}(a) shows ${\overline{\mathrm{SNR}}}$~\eqref{Eq:Mean_c1_c3_c4_s2} as a function of ${\sigma_{\theta_{\mathrm{B}}}}$ and ${\theta_{\mathrm{\mathrm{UE}}}}$. It is observed that, for a given $\sigma_{\theta_{\mathrm{B}}}$ as ${\theta_{\mathrm{\mathrm{UE}}}}$ increases, ${\overline{\mathrm{SNR}}}$ is reduced. This is more evident with the increase of $\sigma_{\theta_{\mathrm{B}}}$. For example, for ${\sigma_{\theta_{\mathrm{B}}}=0.1^o}$ and ${\theta_{\mathrm{UE}}}$ equal to $0^o$ and $60^o$ the resulting ${\overline{\mathrm{SNR}}}$ is $3.714$ and $3.697$, respectively. Meanwhile, by setting ${\sigma_{\theta_{\mathrm{B}}}=6^o}$ and ${\theta_{\mathrm{UE}}}$ equal to $0^o$, $30^o$ and $60^o$ the resulting ${\overline{\mathrm{SNR}}}$ is ${1.905}$, ${1.707}$ and ${1.063}$, respectively. Also, for a given $\theta_{\mathrm{UE}}$ as $\sigma_{\theta_{\mathrm{B}}}$ increases $\overline{\mathrm{SNR}}$ is reduced. This behavior is more prominent for higher values of $\theta_{\mathrm{UE}}$, where the range of $\sigma_{\theta_{\mathrm{B}}}$ that yields high ${\overline{\mathrm{SNR}}}$ is significantly reduced. As an example, for ${\theta_{\mathrm{UE}}=10^o}$ and $\sigma_{\theta_{\mathrm{B}}}$ equal to $0.5^o$ and $6.5^o$ the resulting $\overline{\mathrm{SNR}}$ is $3.679$ and $1.768$, respectively. On the other hand, for ${\theta_{\mathrm{UE}}=60^o}$ and $\sigma_{\theta_{\mathrm{B}}}$ equal to $0.5^o$ and $6.5^o$ the resulting $\overline{\mathrm{SNR}}$ is $3.568$ and $0.987$, respectively. The aforementioned observations made on Fig.~\ref{fig:Fig_4}(a) are verified by the definitions of $\overline{\mathrm{SNR}}$ in~\eqref{Eq:Mean_c1_c3_c4_s2} and by that of the parameters $\alpha$ and $\beta$ in~\eqref{Eq:a_4} and~\eqref{Eq:b_4}, respectively. In more detail, by increasing $\theta_{\mathrm{UE}}$ the parameter $\alpha$ is reduced, whereas $\beta$ is increased. Moreover combined with the increase of $\sigma_{\theta_{\mathrm{B}}}$ the numerator of~\eqref{Eq:Mean_c1_c3_c4_s2} is reduced and the denominator is increased, hence the $\overline{\mathrm{SNR}}$ deteriorates.

Fig.~\ref{fig:Fig_4}(b), illustrates $s_{\widetilde{\mathrm{SNR}}}$~\eqref{Eq:Skewness_Case4} as a function of $\sigma_{\theta_{\mathrm{B}}}$ and $\theta_{\mathrm{UE}}$. The dashed black line denotes the combination of ${\sigma_{\theta_{\mathrm{B}}}}$ and ${\theta_{\mathrm{UE}}}$ values that yield ${s_{\widetilde{\mathrm{SNR}}}=0}$. It is observed that for a given $\theta_{\mathrm{UE}}$ as $\sigma_{\theta_{\mathrm{B}}}$ increases the skewness increases from negative to positive values. This increase of the misalignment severity expressed in terms of $\sigma_{\theta_{\mathrm{B}}}$ indicates that the non-zero amplitudes of ${f_{\widetilde{\mathrm{SNR}}}\left(x\right)}$~\eqref{Eq:f_SNR_c1_c3_c4_s2} are obtained at lower values of the instantaneous SNR with the increase of $\sigma_{\theta_{\mathrm{B}}}$. Moreover, for a given $\sigma_{\theta_{\mathrm{B}}}$ as $\theta_{\mathrm{UE}}$ increases the skewness also increases. In more detail, for a low $\sigma_{\theta_{\mathrm{B}}}$ and increasing $\theta_{\mathrm{UE}}$ the resulting skewness remains negative, which means that the non-zero values of~\eqref{Eq:f_SNR_c1_c3_c4_s2} are still obtained for relatively high instantaneous SNR. As an example for ${\sigma_{\theta_{\mathrm{B}}}=1.5^o}$ and ${\theta_{\mathrm{UE}}=10^o}$ the skewness is $-2$, while for the same $\sigma_{\theta_{\mathrm{B}}}$ and ${\theta_{\mathrm{UE}}=60^o}$ the skewness is $-1.49$. For a given intermediate $\sigma_{\theta_{\mathrm{B}}}$ as $\theta_{\mathrm{UE}}$ increases the skewness increases but it is close to $0$ from both the negative and positive directions. This indicates that the higher non-zero amplitudes of~\eqref{Eq:f_SNR_c1_c3_c4_s2} are obtained for the low and high values of the instantaneous SNR. As an example, for ${\sigma_{\theta_{\mathrm{B}}}=5^o}$ and changing $\theta_{\mathrm{UE}}$ from $5^o$ to $45^o$ the skewness is $-0.32$ and $0.16$, respectively. Also, for a high $\sigma_{\theta_{\mathrm{B}}}$ and increasing ${\theta_{\mathrm{UE}}}$ the skewness is positive. For example, for ${\sigma_{\theta_{\mathrm{B}}}=10^o}$ and changing $\theta_{\mathrm{UE}}$ from ${0^o}$ to ${60^o}$ the skewness is $0.64$ and $1.62$, respectively. This indicates that in the case of severe misalignment the non-zero amplitudes of~\eqref{Eq:f_SNR_c1_c3_c4_s2} are obtained in the low instantaneous SNR regime.
%
%
\begin{figure}
\centering
\includegraphics[width=0.9\columnwidth]{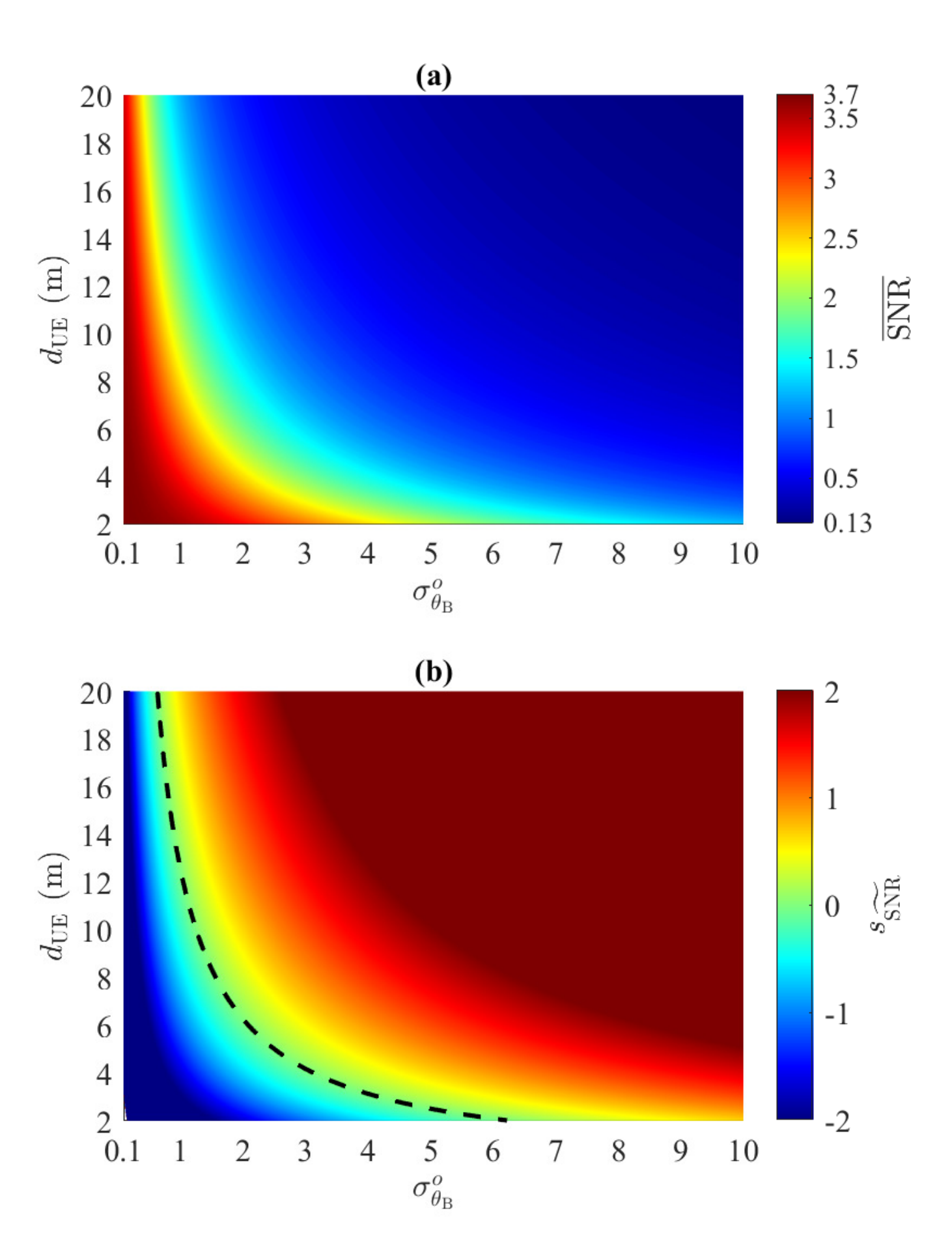}
\caption{Study of misalignment error for variable RIS-UE distance, $d_{\mathrm{UE}}$. (a) $\overline{\mathrm{SNR}}$ and (b) ${s_{\widetilde{\mathrm{SNR}}}}$, as a function of ${\sigma_{\theta_{\mathrm{B}}}}$. The user is along the direction with ${\theta_{\mathrm{UE}}=0^o}$, and the footprint of the incident beam is ${w_{\mathrm{RIS}}=25\text{ cm}}$. The dashed line marks the locus of operation points with ${s_{\widetilde{\mathrm{SNR}}}}=0$.}
\label{fig:Fig_5}
\end{figure}

Fig.~\ref{fig:Fig_5}(a) shows ${\overline{\mathrm{SNR}}}$~\eqref{Eq:Mean_c1_c3_c4_s2} as a function of $\sigma_{\theta_{\mathrm{B}}}$ and $d_{\mathrm{\mathrm{UE}}}$. It is observed that, for a given $d_{\mathrm{\mathrm{UE}}}$ as $\sigma_{\theta_\mathrm{B}}$ increases the $\overline{\mathrm{SNR}}$ is reduced. For example, for ${d_{\mathrm{UE}}=6\text{ m}}$ by changing $\sigma_{\theta_{\mathrm{B}}}$ from $1^o$ to $5^o$ the $\overline{\mathrm{SNR}}$ is $2.85$ and $0.86$, respectively. Moreover, the increase of $d_{\mathrm{UE}}$ reduces the system tolerance to misalignment. As an example, for ${d_{\mathrm{UE}}=2\text{ m}}$ and ${{\sigma_{\theta_{\mathrm{B}}}}=1^o}$, ${\overline{\mathrm{SNR}}=3.58}$, while ${\sigma_{\theta_{\mathrm{B}}}=3^o}$ yields ${\overline{\mathrm{SNR}}=2.85}$. On the other hand, by assuming ${d_{\mathrm{UE}}=20 \text{ m}}$ and changing $\sigma_{\theta_{\mathrm{B}}}$ from $1^o$ to $3^o$ the resulting $\overline{\mathrm{SNR}}$ is $1.23$ and $0.43$, respectively. The previous observations indicate that at high RIS-UE separation distances the link becomes increasingly sensitive to misalignment.

Fig.~\ref{fig:Fig_5}(b), illustrates $s_{\widetilde{\mathrm{SNR}}}$~\eqref{Eq:Skewness_Case4} as a function of $\sigma_{\theta_\mathrm{B}}$ and $d_{\mathrm{\mathrm{UE}}}$. The dashed black line denotes the combination of ${\sigma_{\theta_{\mathrm{B}}}}$ and ${d_{\mathrm{UE}}}$ values that yield ${s_{\widetilde{\mathrm{SNR}}}=0}$. It is observed that for a given $d_{\mathrm{UE}}$ as $\sigma_{\theta_{\mathrm{B}}}$ increases the skewness increases. Also, for a given $\sigma_{\theta_{\mathrm{B}}}$ as $d_{\mathrm{UE}}$ increases the skewness drastically increases. In more detail, the skewness remains negative with the increase of $d_{\mathrm{UE}}$ for a low level of misalignment. For example, for $\sigma_{\theta_{\mathrm{B}}}=0.3^o$ and changing $d_{\mathrm{UE}}$ from $5$ to ${20\text{ m}}$ the skewness is $-2$ and ${-1.08}$, respectively. Meanwhile, for the greater values of $d_{\mathrm{UE}}$ the skewness is closer to $0$ as $\sigma_{\theta_B}$ increases. For example, for ${d_{\mathrm{UE}}=15\text{ m}}$ and changing $\sigma_{\theta_{\mathrm{B}}}$ from ${0.7^o}$ to ${1.2^o}$ the skewness is $-0.29$ and $0.38$, respectively. On the other hand, for low values of $d_{\mathrm{UE}}$ the skewness is closer to $0$ starting from the intermediate values of $\sigma_{\theta_{\mathrm{B}}}$. As an example, for ${d_{\mathrm{UE}}=3\text{ m}}$ and changing $\sigma_{\theta_{\mathrm{B}}}$ from ${4^o}$ to ${5^o}$ the resulting skewness is $-0.07$ and $0.24$, respectively. Furthermore, by increasing $\sigma_{\theta_{\mathrm{B}}}$ beyond its intermediate values, even for a low $d_{\mathrm{UE}}$ the skewness is positive. Hence, the non-zero values of the amplitude of~\eqref{Eq:f_SNR_c1_c3_c4_s2} are found on the low instantaneous SNR range.
\subsection{Qualitative differences between misalignment on the steering plane and the steering plane normal}\label{Sec:Compare_Case4_with_Case3}

For misalignment that takes place on the steering plane normal, the stochastic behavior of the RIS-aided link can be similarly studied if~\eqref{Eq:SNR_case3_approx_s2} is used in place of~\eqref{Eq:SNR_form_c1_c3_c4}. Note, however that while the parameter $\alpha$ given by~\eqref{Eq:a_4} is common for both orientations of misalignment, the parameters $\beta, \zeta$ differ in that $\beta$ of~\eqref{Eq:b_4} depends on $\theta_{UE}$, whereas $\zeta$ in~\eqref{Eq:b_3} does not. As a consequence, the misalignment error here is qualitatively different with respect to the case studied in the previous examples and is associated with the elliptical shape that the beam acquires upon steering. The major axis of the elliptical beam cross-section lies on the steering plane and increases with increasing steering angle, while the minor axis that is oriented along the steering plane normal does not depend on the steering angle (this can be easily deduced from~\eqref{Eq:Srinf}, e.g. by setting ${\phi_{\mathrm{B}}=0}$). This qualitative difference has distinct implications on the link robustness with respect to the two misalignment cases. This becomes clear for the limiting case of large AP beam footprint $w_{\mathrm{RIS}}$ (corresponding to relatively low AP antenna gain), where ${z_R\gg d_{\mathrm{UE}}}$, and~\eqref{Eq:a_4} can be approximated~as
\begin{align}
\alpha \approx \frac{2 P_t \left|R\right|^2 A_r}{N_o \pi w_{\mathrm{\mathrm{RIS}}}^2}.
\label{Eq:alpha_approx}
\end{align}
Also, $\beta$ in~\eqref{Eq:b_4} can be approximated~as
\begin{align}
\beta\approx \frac{k_o d_{\mathrm{UE}}^2}{z_R \cos^2\left({\theta_{\mathrm{UE}}}\right)},
\label{Eq:b_4_approx1}
\end{align}
whereas $\zeta$ in~\eqref{Eq:b_3} can be approximated~as
\begin{align}
\zeta\approx\frac{k_o d_{\mathrm{UE}}^2}{z_R}.
\label{Eq:b_3_approx1}
\end{align}
With simple inspection of~\eqref{Eq:alpha_approx}\text{--}\eqref{Eq:b_3_approx1}, it becomes evident that~\eqref{Eq:SNR_case3_approx_s2} is independent of $\theta_{\mathrm{UE}}$, whereas~\eqref{Eq:SNR_form_c1_c3_c4} is not. Similar conclusions can be deduced for high gain AP beams, where $w_{\mathrm{RIS}}$ is relatively small, leading to ${d_{\mathrm{UE}}\gg z_R}$, and the parameters $\alpha, \beta$ and $\zeta$ can be approximated~as 
\begin{align}
\alpha \approx \frac{2 P_t \left|R\right|^2 A_r}{N_o \pi w_{\mathrm{\mathrm{RIS}}}^2}\frac{z_R^2}{d_{\mathrm{UE}}^2}\cos^2{\left(\theta_{\mathrm{UE}}\right)},
\label{Eq:alpha_approx2}
\end{align}
\begin{align}
\beta\approx k_oz_R\cos^2{\left(\theta_{\mathrm{UE}}\right)}
\label{Eq:b_4_approx2}
\end{align}
and
\begin{align}
\zeta\approx k_o z_R,
\label{Eq:b_3_approx2}
\end{align}
respectively.
\section{Conclusion}\label{Sec:Coclusion}
In this work, beam misalignment in RIS-aided THz links was studied theoretically. To analytically characterize the stochastic performance of the link, a model that treats the RIS as a continuous surface was employed and the RIS was considered to be sufficiently large to capture the entire footprint of the incident beam. The analytical model revealed qualitative differences that are expected in the RIS performance with respect to the orientation of misalignment. It was found that misalignment errors that occur on the steering plane have stronger dependence on the steering angle with respect to misalignment taking place on the perpendicular plane, leading to different robustness of the communication link between the two cases. The analytical models, which were validated with numerical calculations, provide the necessary tools for assessing the stochastic RIS performance with respect to crucial link parameters, such as the transmitter's beam width, the transmitter-RIS distance, the RIS-receiver distance, and the steering angle of the RIS.

\section*{Acknowledgements}
This work has received funding from the European Commission Horizon $2020$ research and innovation programme ARIADNE under grant agreement No. $871464$.

{\appendices

\section*{Appendix A}\label{Sec:Appendix_A}
The trigonometric functions involving the angles $\theta_{\mathrm{B}}, \phi_{\mathrm{B}}$ that appear in~\eqref{Eq:x_B}--\eqref{Eq:z_B} can be expressed in terms of the misalignment error angles $\delta \theta_x, \delta \theta_y$ as:
\begin{align}
\sin\left(\theta_{\mathrm{B}}\right)\cos\left(\phi_{\mathrm{B}}\right) = \sin\left(\theta_{\mathrm{UE}}+\delta\theta_x\right) \cos\left(\delta \theta_y\right),
\label{Eq:EqA1}
\end{align}
\begin{align}
\sin\left(\theta_{\mathrm{B}}\right)\sin\left(\phi_{\mathrm{B}}\right) = \sin\left(\delta \theta_y\right).
\label{Eq:EqA2}
\end{align}
\begin{align}
\cos\left(\theta_{\mathrm{B}}\right) = \cos\left(\theta_{\mathrm{UE}}+\delta\theta_x\right) \cos\left(\delta \theta_y\right).
\label{Eq:EqA3}
\end{align}
When misalignment occurs on the steering plane, then ${\phi_{\mathrm{B}}=0}$, ${\delta\theta_y=0}$ and ${\delta\theta_x \neq 0}$. For relatively small error $\delta\theta_x$, it is found with the aid of \eqref{Eq:EqA1}--\eqref{Eq:EqA3} that:
\begin{align}
\cos\left(\theta_{\mathrm{B}}\right) \approx \cos\left(\theta_{\mathrm{UE}}\right)-\sin\left(\theta_{\mathrm{UE}}\right)\delta\theta_x,
\label{Eq:TRIGa1}
\end{align}
\begin{align}
\sin\left(\theta_{\mathrm{B}}\right) \approx \sin\left(\theta_{\mathrm{UE}}\right)+\cos\left(\theta_{\mathrm{UE}}\right)\delta\theta_x,
\label{Eq:TRIGa2}
\end{align}
\begin{align}
\cos\left(\phi_{\mathrm{B}}\right) = 1,
\label{Eq:TRIGa3}
\end{align}
\begin{align}
\sin\left(\phi_{\mathrm{B}}\right) = 0.
\label{Eq:TRIGa4}
\end{align}
Using ~\eqref{Eq:TRIGa1}--\eqref{Eq:TRIGa4} in~\eqref{Eq:Srinf},~\eqref{Eq:x_B}--\eqref{Eq:z_B} and keeping the lowest order $\delta\theta_x$ terms, leads to the approximation of ~\eqref{Eq:SNR_1}, presented in~\eqref{Eq:SNR_form_c1_c3_c4}. By setting the random variable ${X=\delta\theta_x}$ and by assuming that $x$ is an instance of $X$ it is deducted that~\eqref{Eq:SNR_form_c1_c3_c4} is expressed as a function of $x$. Then, by setting ${y=\widetilde{\mathrm{SNR}}\left(x\right)}$ in~\eqref{Eq:SNR_form_c1_c3_c4} and solving the expression in terms of $x$ the
\begin{align}
x_{1,2}=\pm\sqrt{\frac{\ln\left(\frac{y}{\alpha}\right)}{-\beta}}
\label{Eq:x1_2_c1_c3_c4}
\end{align}
roots are obtained. In order for $x_{1,2}$ to be real ${y\leq\alpha}$. Meanwhile, the absolute value of the first order derivative of~\eqref{Eq:SNR_form_c1_c3_c4} is obtained~as
\begin{align}
\left|g'\left(x\right)\right|=\left|\partial \frac{ \alpha \exp\left(-\beta x^2\right)}{\partial x}\right|=2\alpha \beta \left|x\right|
\exp\left(-\beta x^2\right).
\label{Eq:abs_gx}
\end{align}
The PDF of $X$ is obtained as~\cite{papoulis}
\begin{align}
f_X\left(x\right)=\frac{\exp\left(-\frac{x^2}{2\sigma^2}\right)}{\sqrt{2 \pi}\sigma}.
\label{Eq:PDF_Normal}
\end{align}
Then, according to~\cite[Eq. 5.16]{papoulis} and by employing~\eqref{Eq:x1_2_c1_c3_c4}--\eqref{Eq:PDF_Normal}, the PDF of $\widetilde{\mathrm{SNR}}$ is obtained~as
\begin{align}
f_{\widetilde{\mathrm{SNR}}}\left(y\right)=\frac{f_X\left(x_1\right)}{\left|g'\left(x_1\right)\right|}+\frac{f_X\left(x_2\right)}{\left|g'\left(x_2\right)\right|},
\label{Eq:f_SNR_c1_c3_c4_s1}
\end{align}
which after some algebraic manipulations can be equivalently rewritten as~\eqref{Eq:f_SNR_c1_c3_c4_s2}, by setting ${\sigma=\sigma_{\theta_{\mathrm{B}}}}$. Then, by employing~\eqref{Eq:f_SNR_c1_c3_c4_s2}, the CDF of $\widetilde{\mathrm{SNR}}$ is defined~as
\begin{align}
F_{\widetilde{\mathrm{SNR}}}\left(z\right)=\int _{0}^{z} \frac{\left(\frac{y}{\alpha}\right)^{\frac{1}{2 \beta \sigma^2}}}{\sqrt{2\pi}\sigma y \beta
\sqrt{\frac{\left|\ln\left(\frac{y}{\alpha}\right)\right|}{\beta}}}\mathrm{d}y,
\label{Eq:F_SNR_c1_c3_c4_s1}
\end{align}
which after some algebraic manipulations can be equivalently rewritten as~\eqref{Eq:F_SNR_c1_c3_c4_s2}, by setting ${\sigma=\sigma_{\theta_{\mathrm{B}}}}$ . The mean of the random variable $\widetilde{\mathrm{SNR}}$ with the aid of~\eqref{Eq:f_SNR_c1_c3_c4_s2} is defined as~\cite[Eq. (5.67)]{papoulis}
\begin{align}
\overline{\mathrm{SNR}}=\int _{0}^{a} y 
\frac{\left(\frac{y}{\alpha}\right)^{\frac{1}{2 \beta \sigma^2}}}{\sqrt{2\pi}\sigma y \beta
\sqrt{\frac{\left|\left(\frac{y}{\alpha}\right)\right|}{\beta}}}
\mathrm{d}y,
\label{Eq:Mean_c1_c3_c4_s1}
\end{align}
which after some algebraic manipulations can evaluated as~\eqref{Eq:Mean_c1_c3_c4_s2}, by setting ${\sigma=\sigma_{\theta_{\mathrm{B}}}}$. The variance of $\widetilde{\mathrm{SNR}}$ is defined as~\cite[Eq. (5.68)]{papoulis}
\begin{align}
\sigma^2_{\widetilde{\mathrm{SNR}}}=\mathbb{E}\left[\widetilde{\mathrm{SNR}}^2\right]-\left(\mathbb{E}\left[\widetilde{\mathrm{SNR}}\right]\right)^2.
\label{Eq:Var_c1_c3_c4_s1}
\end{align}
By employing~\eqref{Eq:f_SNR_c1_c3_c4_s2} and~\eqref{Eq:Mean_c1_c3_c4_s2}, $\sigma^2_{\widetilde{\mathrm{SNR}}}$ in~\eqref{Eq:Var_c1_c3_c4_s1} can be equivalently rewritten~as
\begin{align}
\begin{split}
\sigma_{\widetilde{\mathrm{SNR}}}^2&=\int _{0}^{a}
y^2
\frac{\left(\frac{y}{\alpha}\right)^{\frac{1}{2 \beta \sigma^2}}}{\sqrt{2\pi}\sigma y \beta
\sqrt{\frac{\left|\left(\frac{y}{\alpha}\right)\right|}{\beta}}}
\mathrm{d}y \\
& -
\left(\frac{\alpha}{\sqrt{1+2\beta\sigma^2}}\right)^2,
\end{split}
\label{Eq:Var_c1_c3_c4_s2}
\end{align}
which after some algebraic manipulations can be evaluated as~\eqref{Eq:Var_c1_c3_c4_s3}, by setting ${\sigma=\sigma_{\theta_{\mathrm{B}}}}$. The skewness of ${\widetilde{\mathrm{SNR}}}$ is defined~as~\cite[Eq. (5.70)]{papoulis}
\begin{align}
s_{\widetilde{\mathrm{SNR}}}=\frac{\mathbb{E}\left[\widetilde{\mathrm{SNR}}^3\right]-3\mathbb{E}\left[\widetilde{\mathrm{SNR}}\right]\mathbb{V}\left[\widetilde{\mathrm{SNR}}\right]-\left(\mathbb{E}\left[\widetilde{\mathrm{SNR}}\right]\right)^3
}{\left(\sqrt{\mathbb{V}\left[\widetilde{\mathrm{SNR}}\right]}\right)^3},
\label{Eq:Skewness_Case4_s1}
\end{align}
where
\begin{align}
\mathbb{E}\left[\widetilde{\mathrm{SNR}}^3\right]=\int _{0}^{\alpha} y^2 \frac{\left(\frac{y}{\alpha}\right)^{\frac{1}{2\beta\sigma^2}}}
{\sqrt{2 \pi}\sigma \beta\sqrt{\frac{\left|\ln\left(\frac{y}{\alpha}\right)\right|}{\beta}}}
\mathrm{d}y,
\end{align}
which is equivalently rewritten~as
\begin{align}
\mathbb{E}\left[\widetilde{\mathrm{SNR}}^3\right]=\frac{\alpha^3}{\sqrt{1+6 \beta \sigma^2}}.
\label{Eq:mean_to_3}
\end{align}
Then,~\eqref{Eq:Skewness_Case4_s1} by employing~\eqref{Eq:mean_to_3},~\eqref{Eq:Mean_c1_c3_c4_s2} and~\eqref{Eq:Var_c1_c3_c4_s3} and after some algebraic manipulations can be equivalently rewritten as~\eqref{Eq:Skewness_Case4}, by setting ${\sigma=\sigma_{\theta_x}}$.
\section*{Appendix B}\label{Sec:Appendix_B}
When misalignment occurs on the steering plane normal, then ${\phi_{\mathrm{B}}\neq 0}$, ${\delta\theta_x=0}$ and ${\delta\theta_y \neq 0}$. For relatively small error $\delta\theta_y$, it is found with the aid of \eqref{Eq:EqA1}--\eqref{Eq:EqA3} that:
\begin{align}
\cos\left(\theta_{\mathrm{B}}\right) \approx \cos\left(\theta_{\mathrm{UE}}\right),
\label{Eq:TRIGb1}
\end{align}
\begin{align}
\sin\left(\theta_{\mathrm{B}}\right) \approx \sin\left(\theta_{\mathrm{UE}}\right),
\label{Eq:TRIGb2}
\end{align}
\begin{align}
\cos\left(\phi_{\mathrm{B}}\right) \approx 1,
\label{Eq:TRIGb3}
\end{align}
\begin{align}
\sin\left(\phi_{\mathrm{B}}\right) \approx \frac{\delta\theta_y}{\sin\left(\theta_{\mathrm{UE}}\right)}.
\label{Eq:TRIGb4}
\end{align}
By replacing~\eqref{Eq:TRIGb1}--\eqref{Eq:TRIGb4} in~\eqref{Eq:Srinf},~\eqref{Eq:x_B}--\eqref{Eq:z_B} and keeping the lowest order $\delta\theta_y$ terms, the approximation of the SNR expression~\eqref{Eq:SNR_1} yields~\eqref{Eq:SNR_case3_approx_s2}. Then, by setting the Gaussian random variable ${X=\delta\theta_y}$, where ${X \sim N\left(0,\sigma^2\right)}$ and by assuming that $x$ is an instance of $X$ it is deducted that~\eqref{Eq:SNR_case3_approx_s2} is expressed as a function of $x$. Then, the proof of the PDF, CDF, average SNR, variance and skewness of~\eqref{Eq:f_SNR_c1_c3_c4_s2}--\eqref{Eq:Skewness_Case4_s1} can be obtained as in Appendix A, by replacing $\beta$~\eqref{Eq:b_4} with $\zeta$~\eqref{Eq:b_3} and by setting ${\sigma=\sigma_{\theta_y}}$.
}

\bibliographystyle{IEEEtran}
\bibliography{References}

\begin{thebibliography}{10}
\providecommand{\url}[1]{#1}
\csname url@samestyle\endcsname
\providecommand{\newblock}{\relax}
\providecommand{\bibinfo}[2]{#2}
\providecommand{\BIBentrySTDinterwordspacing}{\spaceskip=0pt\relax}
\providecommand{\BIBentryALTinterwordstretchfactor}{4}
\providecommand{\BIBentryALTinterwordspacing}{\spaceskip=\fontdimen2\font plus
\BIBentryALTinterwordstretchfactor\fontdimen3\font minus
  \fontdimen4\font\relax}
\providecommand{\BIBforeignlanguage}[2]{{%
\expandafter\ifx\csname l@#1\endcsname\relax
\typeout{** WARNING: IEEEtran.bst: No hyphenation pattern has been}%
\typeout{** loaded for the language `#1'. Using the pattern for}%
\typeout{** the default language instead.}%
\else
\language=\csname l@#1\endcsname
\fi
#2}}
\providecommand{\BIBdecl}{\relax}
\BIBdecl

\bibitem{A:Roadmap_to_6G}
K.~B. Letaief, W.~Chen, Y.~Shi, J.~Zhang, and Y.-J.~A. Zhang, ``{T}he {R}oadmap
  to 6{G}: {AI} {E}mpowered {W}ireless {N}etworks,'' \emph{IEEE Commun. Mag.},
  vol.~57, no.~8, pp. 84--90, Aug. 2019.

\bibitem{B:Next_Gen_Wir_THz_Com_Netw}
A.-A.~A. Boulogeorgos and A.~Alexiou, \emph{{N}ext Generation Wireless
  {T}erahertz Communication Networks}.\hskip 1em plus 0.5em minus 0.4em\relax
  CRC Press, Sep. 2020, ch. {A}ntenna misalignment and blockage in {T}Hz
  communications.

\bibitem{A:amakawa2021white}
S.~Amakawa, Z.~Aslam, J.~Buckwater, S.~Caputo, A.~Chaoub, Y.~Chen, Y.~Corre,
  M.~Fujishima, Y.~Ganghua, S.~Gao \emph{et~al.}, ``White {P}aper on {RF}
  {E}nabling 6{G}--{O}pportunities and {C}hallenges from {T}echnology to
  {S}pectrum,'' \emph{Res. Vis.}, vol.~13, pp. 1--68, Sep. 2021.

\bibitem{Me_Sc_Reps_21}
E.~N. Papasotiriou, A.-A.~A. Boulogeorgos, K.~Haneda, M.~F. de~Guzman, and
  A.~Alexiou, ``An experimentally validated fading model for {TH}z wireless
  systems,'' \emph{Sc. Rep.}, vol.~11, no.~1, pp. 1--14, Sep. 2021.

\bibitem{jornet2011}
J.~M. Jornet and I.~F. Akyildiz, ``Channel modeling and capacity analysis for
  electromagnetic wireless nanonetworks in the terahertz band,'' \emph{IEEE
  Trans. Wir. Commun.}, vol.~10, no.~10, pp. 3211--3221, Oct. 2011.

\bibitem{J:Kokkoniemi_100_450GHz}
J.~Kokkoniemi, L.~Janne, and J.~Markku, ``A line-of-sight channel model for the
  100–450 gigahertz frequency band,'' \emph{EURASIP J. on Wir. Commun. and
  Net.}, no.~88, Apr. 2021.

\bibitem{Mag:RIS_pathl_alex}
A.~A.~A. {Boulogeorgos} and A.~{Alexiou}, ``{C}overage {A}nalysis of
  {R}econfigurable {I}ntelligent {S}urface {A}ssisted {TH}z {W}ireless
  {S}ystems,'' \emph{IEEE Op. J. of Veh. Techn.}, vol.~2, pp. 94--110, Jan.
  2021.

\bibitem{Mag:Analytical_Performance_Assessment_of_THz_Wireless_Systems}
A.-A.~A. {Boulogeorgos}, E.~N. {Papasotiriou}, and A.~{Alexiou}, ``Analytical
  performance assessment of {TH}z wireless systems,'' \emph{IEEE Access},
  vol.~7, pp. 11\,436--11\,453, Jan. 2019.

\bibitem{A:On_millimeter_wave_and_THz_mobile_radio_channel_for_smart_rail_mobility}
K.~Guan, G.~Li, T.~Kürner, A.~F. Molisch, B.~Peng, R.~He, B.~Hui, J.~Kim, and
  Z.~Zhong, ``On millimeter wave and {THz} mobile radio channel for smart rail
  mobility,'' \emph{IEEE Trans. Veh. Technol.}, vol.~66, no.~7, pp. 5658--5674,
  Jul. 2017.

\bibitem{C:EGW_Imp_Beam_Misal_RIS_assist_THz_Sys}
E.~N. Papasotiriou, A.-A.~A. Boulogeorgos, and A.~Alexiou, ``On the {I}mpact of
  {B}eam {M}isalignment in {R}econfigurable {I}ntelligent {S}urface {A}ssisted
  {TH}z {S}ystems,'' in \emph{2021 IEEE 22nd Int. Work. on Sig. Proc. Adv. in
  Wir. Commun. (SPAWC)}, Sep. 2021, pp. 121--125.

\bibitem{A:Dir_THz_Commun_Sys_for_6G_Fact_Check}
A.-A.~A. Boulogeorgos, J.~M. Jornet, and A.~Alexiou, ``Directional {T}erahertz
  {C}ommunication {S}ystems for 6{G}: {F}act {C}heck,'' \emph{IEEE Veh. Techn.
  Mag.}, vol.~16, no.~4, pp. 68--77, Dec. 2021.

\bibitem{C:PIMRC_2019_turkey}
A.-A.~A. {Boulogeorgos}, E.~N. {Papasotiriou}, and A.~{Alexiou}, ``Analytical
  performance evaluation of {TH}z wireless fiber extenders,'' in \emph{IEEE
  30th An. Int. Symp. on Per., Indoor and Mob. Radio Commun. (PIMRC)},
  Istanbul, Turkey, Sep. 2019, pp. 1--6.

\bibitem{Mag:Perf_Anal_of_THz_Wir_Sys_in_the_Pres_of_Ant_Misal_and_PHN}
E.~N. {Papasotiriou}, A.-A.~A. {Boulogeorgos}, and A.~{Alexiou}, ``Performance
  {A}nalysis of {THz} {W}ireless {S}ystems in the {P}resence of {A}ntenna
  {M}isalignment and {P}hase {N}oise,'' \emph{IEEE Commun. Let.}, Mar. 2020.

\bibitem{C:Relay-Based_Blockage_and_Antenna_Misalignment_Mitigation_in_THz_Wireless_Communications}
G.~{Stratidakis}, E.~N. {Papasotiriou}, H.~{Konstantinis}, A.-A.~A.
  {Boulogeorgos}, and A.~{Alexiou}, ``Relay-based blockage and antenna
  misalignment mitigation in {THz} wireless communications,'' in \emph{2nd 6G
  Wir. Sum. (6G SUMMIT)}, Mar. 2020, pp. 1--4.

\bibitem{A:Joint_Imp_Ph_err_HD_imp_mob_Int_RIS_over_kappa_mu}
A.~Sikri, A.~Mathur, and G.~Kaddoum, ``Joint {I}mpact of {P}hase {E}rror,
  {T}ransceiver {H}ardware {I}mpairments, and {M}obile {I}nterferers on
  {RIS}-{A}ided {W}ireless {S}ystem {O}ver $\kappa-\mu$ {F}ading {C}hannels,''
  \emph{IEEE Commun. Let.}, vol.~26, no.~10, pp. 2312--2316, Jul. 2022.

\bibitem{C:Eff_Rate_of_RIS_Netw_Loc_and_Ph_Est_Unc}
L.~Kong, S.~Kisseleff, S.~Chatzinotas, B.~Ottersten, and M.~Erol-Kantarci,
  ``Effective {R}ate of {RIS}-aided {N}etworks with {L}ocation and {P}hase
  {E}stimation {U}ncertainty,'' in \emph{2022 IEEE Wir. Commun. and Netw. Conf.
  (WCNC)}, Apr. 2022, pp. 2071--2075.

\bibitem{C:OP_Anal_of_RIS_Wir_Netw_W_Von_Mises_Ph_Er}
T.~Wang, M.-A. Badiu, G.~Chen, and J.~P. Coon, ``Outage {P}robability
  {A}nalysis of {R}is-{A}ssisted {W}ireless {N}etworks {W}ith {V}on {M}ises
  {P}hase {E}rrors,'' \emph{IEEE Wireless Communications Letters}, vol.~10,
  no.~12, pp. 2737--2741, Dec. 2021.

\bibitem{C:mmWave_Human_Blockage_at_73_GHz_with_a_Simple_Double_Knife_Edge_Diffraction_Model_and_Extension_for_Directional_Antennas}
G.~R. {MacCartney}, S.~{Deng}, S.~{Sun}, and T.~S. {Rappaport},
  ``Millimeter-wave human blockage at 73 {GHz} with a simple double knife-edge
  diffraction model and extension for directional antennas,'' in \emph{2016
  IEEE 84th Veh. Techn. Conf. (VTC-Fall)}, 2016, pp. 1--6.

\bibitem{C:Frequency_domain_scattering_loss_in_THz_band}
J.~Kokkoniemi, J.~Lehtom\"aki, and M.~Juntti, ``Frequency domain scattering
  loss in {THz} band,'' in \emph{Global Symposium on Millimeter-Waves (GSMM)},
  Montreal, Canada, May 2015, pp. 1--3.

\bibitem{Mag:Beamf_Eff_RIS_giorgos_2021}
G.~Stratidakis, S.~Droulias, and A.~Alexiou, ``Analytical {P}erformance
  {A}ssessment of {B}eamforming {E}fficiency in {R}econfigurable {I}ntelligent
  {S}urface-{A}ided {L}inks,'' \emph{IEEE Access}, vol.~9, pp.
  115\,922--115\,931, Aug. 2021.

\bibitem{C:RIS_eff_giorgos_2022}
------, ``Understanding the {RIS} efficiency: from partial to full
  illumination,'' in \emph{2022 IEEE 23rd Int. Workshop on Sig. Proc. Adv. in
  Wi. Commun. (SPAWC)}, Jul. 2022, pp. 1--5.

\bibitem{C:Impact_RIS_size_beam_eff_giorgos_2021}
------, ``Impact of {R}econfigurable {I}ntelligent {S}urface size on
  beamforming efficiency,'' in \emph{2021 IEEE 32nd An. Int. Symp. on Per.,
  Ind. and Mob. Radio Commun. (PIMRC)}, 2021, pp. 1--5.

\bibitem{A:Giorgos_Opt_pos_orient_RIS_mob_user}
------, ``Optimal position and orientation study of {R}econfigurable
  {I}ntelligent {S}urfaces in a mobile user environment,'' \emph{IEEE Trans. on
  Ant. and Prop.}, pp. 1--1, Sep. 2022.

\bibitem{stratidakis2021analytical}
------, ``An analytical framework for {R}econfigurable {I}ntelligent {S}urfaces
  placement in a mobile user environment,'' in \emph{Proc. of the 19th ACM
  Conf. on Emb. Netw. Sensor Sys.}, Nov. 2021, pp. 623--627.

\bibitem{BADARNEH2022}
\BIBentryALTinterwordspacing
O.~S. Badarneh, R.~Mesleh, and Y.~M. Khattabi, ``Reconfigurable intelligent
  surfaces-assisted terahertz communications,'' \emph{J. of the Franklin
  Inst.}, Jul. 2022. [Online]. Available:
  \url{https://www.sciencedirect.com/science/article/pii/S0016003222004434}
\BIBentrySTDinterwordspacing

\bibitem{A:Alx_Casc_Comp_Turb_Misal_RIS}
A.-A.~A. Boulogeorgos, N.~D. Chatzidiamantis, H.~G. Sandalidis, A.~Alexiou, and
  M.~D. Renzo, ``Cascaded {C}omposite {T}urbulence and {M}isalignment:
  {S}tatistical {C}haracterization and {A}pplications to {R}econfigurable
  {I}ntelligent {S}urface-{E}mpowered {W}ireless {S}ystems,'' \emph{IEEE Trans.
  on Veh. Techn.}, vol.~71, no.~4, pp. 3821--3836, Apr. 2022.

\bibitem{A:Ex_anal_RIS_aided_THz_over_a_m_point_errors}
V.~K. Chapala and S.~M. Zafaruddin, ``Exact {A}nalysis of {RIS}-{A}ided {TH}z
  {W}ireless {S}ystems {O}ver $\alpha-\mu$ {F}ading {W}ith {P}ointing
  {E}rrors,'' \emph{IEEE Commun. Let.}, vol.~25, no.~11, pp. 3508--3512, Nov.
  2021.

\bibitem{2020arXiv201200267D}
H.~Du, J.~Zhang, K.~Guan, D.~Niyato, H.~Jiao, Z.~Wang, and T.~Kürner,
  ``Performance and {O}ptimization of {R}econfigurable {I}ntelligent {S}urface
  {A}ided {TH}z {C}ommunications,'' \emph{IEEE Trans. on Commun.}, vol.~70,
  no.~5, pp. 3575--3593, May 2022.

\bibitem{A:Alx_OP_RIS_UAV_Misal}
A.-A. A.~Boulogeorgos, A.~Alexiou, and M.~D. Renzo, ``Outage performance
  analysis of {RIS}-assisted {UAV} wireless systems under disorientation and
  misalignment,'' \emph{IEEE Trans. on Veh. Techn.}, pp. 1--16, Jun. 2022.

\bibitem{C:Alx_Perf_Anal_Multi_RIS_Surf_Emp_THz_Wir_Sys}
A.-A.~A. Boulogeorgos, N.~Chatzidiamantis, H.~G. Sandalidis, A.~Alexiou, and
  M.~Di~Renzo, ``Performance {A}nalysis of {M}ulti-{R}econfigurable
  {I}ntelligent {S}urface-{E}mpowered {TH}z {W}ireless {S}ystems,'' in
  \emph{ICC 2022 - IEEE Int. Conf. on Commun.}, May 2022, pp. 1481--1487.

\bibitem{papoulis}
A.~Papoulis and S.~Pillai, \emph{Probability, Random Variables, and Stochastic
  Processes}, ser. McGraw-Hill series in electrical engineering: Commun. and
  signal processing.\hskip 1em plus 0.5em minus 0.4em\relax Tata McGraw-Hill,
  2002.

\bibitem{A:RIS_Disc_Ph_Shift}
P.~Xu, G.~Chen, Z.~Yang, and M.~D. Renzo, ``Reconfigurable {I}ntelligent
  {S}urfaces-{A}ssisted {C}ommunications {W}ith {D}iscrete {P}hase {S}hifts:
  {H}ow {M}any {Q}uantization {L}evels {A}re {R}equired to {A}chieve {F}ull
  {D}iversity?'' \emph{IEEE Wir. Commun. Let.}, vol.~10, no.~2, pp. 358--362,
  Feb. 2021.

\bibitem{A:Beam_Opt_RIS_Disc_Ph_Shits}
Q.~Wu and R.~Zhang, ``Beamforming {O}ptimization for {W}ireless {N}etwork
  {A}ided by {I}ntelligent {R}eflecting {S}urface {W}ith {D}iscrete {P}hase
  {S}hifts,'' \emph{IEEE Trans. on Commun.}, vol.~68, no.~3, pp. 1838--1851,
  Dec. 2020.

\bibitem{Mag:High_Gain_D_Band_Transmitarrays}
F.~{Foglia Manzillo}, A.~{Clemente}, and J.~L. {González-Jiménez},
  ``High-gain {D}-band transmitarrays in standard {PCB} technology for
  beyond-5{G} communications,'' \emph{IEEE Trans. on Ant. and Prop.}, vol.~68,
  no.~1, pp. 587--592, Jan. 2020.

\bibitem{Mag:D_Band_H_Gain_Circ_Pol_Plate_Array_Ant}
M.~M. {Zhou} and Y.~J. {Cheng}, ``{D}-band high-gain circular-polarized plate
  array antenna,'' \emph{IEEE Trans. on Ant. and Prop.}, vol.~66, no.~3, pp.
  1280--1287, Jan. 2018.

\bibitem{C:Two_Types_of_High_Gain_Slot_Array_Ant_based_on_Ridge_Gap_Waveguide_D_Band}
J.~{Liu}, A.~U. {Zaman}, and J.~{Yang}, ``Two types of high gain slot array
  antennas based on ridge gap waveguide in the d-band,'' in \emph{2019 IEEE-APS
  Top. Conf. on Ant. and Prop. in Wir. Commun. (APWC)}, Oct. 2019, pp.
  075--078.

\end{thebibliography}

\end{document}